\documentstyle[12pt, axodraw]{article}
\newcommand{\bc} {\begin{center}}
\newcommand{\ec} {\end{center}}
\newcommand{\be} {\begin{equation}}
\newcommand{\ee} {\end{equation}}
\makeatletter
\def\appendix{\par\clearpage
  \setcounter{section}{0}
  \setcounter{subsection}{0}
  \@addtoreset{equation}{section}
  \def\@sectname{Appendix~}
  \def\theequation{\thesection.\arabic{equation}}
  \def\thesection{\Alph{section}}}
\makeatother

\renewcommand{\theequation}{\thesection.\arabic{equation}}

\begin{document}

\begin{titlepage}
\hskip 11cm \vbox{ \hbox{Budker INP 2003-47} \hbox{DESY-03-083}
\hbox{DFCAL-TH 03/4} \hbox{July 2003}} \vskip 0.3cm

\centerline{\bf THE BOOTSTRAP CONDITIONS FOR THE GLUON REGGEIZATION$^{~\ast}$}

\vskip 0.8cm
\centerline{J.~Bartels$^{a~\dag}$, V.S.~Fadin$^{b~\ddag}$,
R.~Fiore$^{c~\dag\dag}$}

\vskip .3cm \centerline{\sl $^a$ II. Institut f$\ddot{u}$r
Theoretische Physik, Universit\"at Hamburg, Germany}
\centerline{\sl $^{b}$ Institute for Nuclear Physics, 630090
Novosibirsk, Russia} \centerline{\sl and Novosibirsk State
University, 630090 Novosibirsk, Russia} \centerline{\sl $^c$
Dipartimento di Fisica, Universit\`a della Calabria}
\centerline{\sl and Istituto Nazionale di Fisica Nucleare, Gruppo
collegato di Cosenza,} \centerline{\sl I-87036 Arcavacata di
Rende, Cosenza, Italy} \vskip 1cm

\begin{abstract}
Compatibility of gluon Reggeization with $s$-channel unitarity
requires the vertices of the Reggeon interactions to satisfy a
series of  bootstrap conditions. In order to derive, in the
next-to-leading order (NLO), conditions related to  the gluon
production amplitudes, we calculate the $s$-channel
discontinuities of these amplitudes and compare them with those
required by the Reggeization. It turns out that these conditions
include the so called strong bootstrap conditions for the kernel
and for the impact factors of scattering particles, which were
proposed earlier without derivation, and recently were proved to
be satisfied. Besides this, there is a new bootstrap condition,
which relates a number of Reggeon vertices and the gluon
trajectory.

\end{abstract}
\vfill

\hrule \vskip.3cm

\noindent $^{\ast}${\it Work supported in part by the Ministero
dell'Istruzione, dell'Universit\`a e della Ricerca, in part by INTAS and
in part by the Russian Fund of Basic Researches.}

\vfill $ \begin{array}{ll} ^{\dag}\mbox{{\it
e-mail address:}} &
\mbox{BARTELS@MAIL.DESY.DE}\\
^{\ddag}\mbox{{\it
e-mail address:}} &
\mbox{FADIN@INP.NSK.SU}\\
^{\dag\dag}\mbox{{\it e-mail address:}} &
\mbox{FIORE@CS.INFN.IT}\\
\end{array}
$
\vfill

\end{titlepage}
\eject

\section{Introduction}
\setcounter{equation}{0}
The gluon Reggeization~\cite{GST,L76} is one of the remarkable properties
of {QCD},  very important at high energy. In particular, the BFKL
approach~\cite{BFKL} to the description of QCD processes at large
center of mass energy $\sqrt s$ and fixed (not increasing with
$s$) momentum transfer {$\sqrt{-t}$}, {$\,\,s \gg |t|$}, is based
on this property. Let us specify that, when using the notion of `gluon
Reggeization', we mean the Reggeized form which in the
BFKL approach was assumed to be valid for amplitudes with colour octet
quantum number and negative signature in exchange channels with fixed
momentum transfer.
In this paper only such amplitudes will be considered, even if it is
not indicated directly. The assumed form for them will be presented
explicitly in the next Section.

In the leading logarithmic approximation (LLA), where only the
leading terms ($~\alpha _S\ln s)^n$ are resummed, the assumption
concerns the amplitudes in the multi-Regge kinematics (MRK), i.e.
at  large (growing with $s$) invariant masses of any pair of
produced particles and fixed transverse momenta. The Reggeized
form of these amplitudes was proved~\cite{BLF}, so that in the LLA
the BFKL approach has a firm ground.

Now the approach is intensively developed in  the next-to-leading
approximation (NLA), when also the terms $~\alpha _S(\alpha _S\ln
s)^n$ are resummed (for references see, e.g.~\cite{FF98,FP02}). In
these calculations it is assumed that  the Reggeized form of the
MRK amplitudes remains valid in the NLA (with the gluon Regge
trajectory and the Reggeon vertices taken in the NLO). Besides
this, the assumption is extended to production amplitudes in the
quasi-multi-Regge kinematics (QMRK), where a pair of produced
particles  has a fixed (not growing with $s$) invariant mass.

The hypothesis of the gluon Reggeization is extremely strong, even
in the LLA, since amplitudes with any number of produced particles
are expressed in terms of the gluon Regge trajectory and a small
number of Reggeon vertices. It seems very hard to combine the
gluon Reggeization with $s$-channel unitarity. Indeed, comparison
of the amplitudes themselves with their discontinuities in
invariant masses of subsets of produced particles, calculated with
the help of $s$-channel unitarity, gives  an infinite set of
``bootstrap" relations. Nevertheless, it turns out that all of
them can be fulfilled  if the vertices and trajectory satisfy
several bootstrap conditions.  This fact is highly non trivial.
Fulfillment of all these conditions generates the basis on which
the proof of Reggeization in the LLA was constructed~\cite{BLF}.
An analogous proof can be constructed in the NLA as
well~\cite{tbp}. The first step in the proof is to derive all
bootstrap conditions; in the next step is has to be shown that
these conditions are indeed satisfied.

The NLO bootstrap conditions imposed by the bootstrap relations
for elastic amplitudes in the NLA were derived several yeas
ago~\cite{FF98}, and they were shown to be satisfied
(see~\cite{FP02} for references). Recently the conditions
following from the bootstrap requirement for the QMRK amplitudes
were obtained, and their fulfillment was shown in~\cite{FKR}. Note
that, since  the QMRK  in the unitarity relations leads to loss of
large logarithms, the QMRK amplitudes used in the BFKL approach
are expressed through the gluon trajectory and the Reggeon
vertices  taken in the leading order (LO).

In this paper we  derive the  NLO bootstrap conditions imposed by
the requirement of  the  Reggeized form of the gluon production
amplitudes  in the MRK.  We show that these conditions include the
so called strong  bootstrap conditions for the kernel and the
impact factors of scattering particles, which were proposed
earlier~\cite{B99,BV00} without derivation, and recently were
proved to be satisfied~\cite{FP02}. Besides this, there is a new
bootstrap condition, which entangles a number of the Reggeon
vertices and the gluon trajectory.

The derivation is based on the calculation of  $s$-channel
discontinuities of the production amplitudes. It is shown that
certain combinations of discontinuities in the NLA can be expressed through
partial derivatives of the real parts of the amplitudes with respect to
subenergies of produced particles. Starting from Reggeization these
derivatives are found to have a well-defined form. By deriving these
combinations of energy discontinuities from the calculated unitarity integrals,
and by then comparing them with the form required by the Reggeization
we obtain the bootstrap conditions.

The paper is organized in the following way. In the next Section
the necessary definitions and notations are introduced, and the
multi-Regge form of QCD amplitudes is presented. In Section 3 we
find the connection between  the $s$-channel discontinuities of
production amplitudes and the partial derivatives of these
amplitudes. Assuming the gluon Reggeization, this connection gives
us the bootstrap relations which express the discontinuities through the
amplitudes themselves. Section 4 is devoted to the calculation of
the discontinuities. The bootstrap conditions on the Reggeon
vertices and trajectory imposed by the bootstrap relations are
derived in Section 5.  Section 6 summarizes the obtained results.

Throughout the paper we work in the NLA, and all equations below
should not be understood with higher accuracy.

\section{Multi-Regge form of QCD amplitudes}

\setcounter{equation}{0}

To specify the hypothesis of the gluon Reggeization in the form,
which was used in the BFKL approach, we briefly review the general
structure of the MRK amplitudes.  Multi-particle amplitudes depend
upon more than one energy variables. Since the general bootstrap
relations, which we will derive and discuss in this paper, involve
discontinuities not only in the total energy but also in
subenergies, we have to discuss the full analytic structure.  At
first sight, this structure appears somewhat complicated, but we
will show that, in the color octet exchange channel in both LA and
NLA, there are substantial simplifications. In particular, there
are combinations of single discontinuities for which the
expressions become simple.

We introduce the light cone momenta $p_1$ and $p_2$ related to
momenta $p_A$ and $p_B$  of colliding particles $A$ and $B$: \be
 p_A=p_1+\left(m_A^2/s\right) p_2~, \;\;
 p_B=p_2+\left(m_B^2/s\right)p_1~, \;\; s=2p_1p_2\simeq (p_A+p_B)^2~,
\ee where $s$ is supposed tending to infinity, and  use the
Sudakov decomposition of momenta  in the form
\begin{equation}
p=\beta  p_1+\alpha p_2 +p_{\perp }~, \,\,\,\,s\alpha \beta =p^2
-p_{\perp }^2=p^2 +\vec p^{~2}~, \label{sud}
\end{equation}
so that the vector sign is used for components of momenta
transverse to the $(p_A, p_B)$ plane. The transverse components
are supposed to be limited (not growing with $s$).

Let us consider production of $n$ particles in the process  $A + B
\;\; \rightarrow \;\; \tilde A + P_1+...+\tilde B$ in the MRK (see
Fig.1). We admit all particles to have non zero masses, reserving
the possibility to consider each of them as a compound state or as
a group of particles. Of course, since we work here in QCD
perturbation theory, our particles are actually partons, i.e.
quarks and gluons.

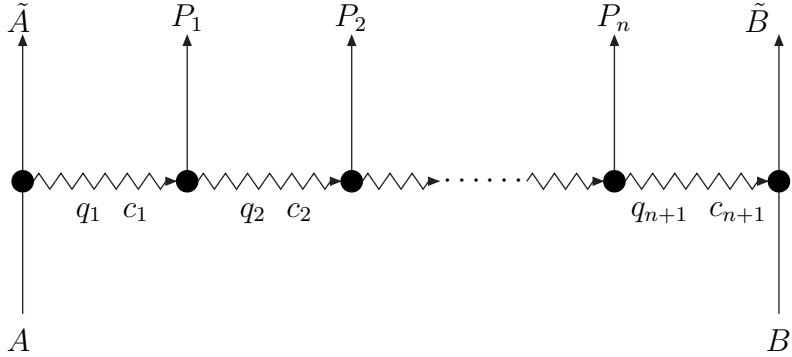
\begin{figure}[htb]
\begin{center}
\begin{picture}(200,120)(35,-80)

\LongArrow(-8,-2)(-8,48) \ZigZag(-4,-6)(46,-6){3}{6}
\ArrowLine(46,-6)(50,-6) \GCirc(-8,-6){4}{0} \Line(-8,-56)(-8,-10)
\Text(-14,56)[l]{$\tilde A$} \Text(-14,-66)[l]{$A$}
\Text(12,-18)[l]{$q_1~~c_1$}

\GCirc(54,-6){4}{0} \LongArrow(54,-2)(54,48)
\ZigZag(58,-6)(108,-6){3}{6} \ArrowLine(108,-6)(112,-6)
\Text(48,56)[l]{$P_1$} \Text(74,-18)[l]{$q_2~~c_2$}

\GCirc(116,-6){4}{0} \LongArrow(116,-2)(116,48)
\ZigZag(120,-6)(145,-6){3}{3} \ArrowLine(145,-6)(149,-6)
\Text(151,-6)[l]{$\cdots \cdots$} \ZigZag(182,-6)(207,-6){3}{3}
\ArrowLine(207,-6)(211,-6) \Text(110,56)[l]{$P_2$}

\GCirc(215,-6){4}{0} \LongArrow(215,-2)(215,48)
\ZigZag(219,-6)(269,-6){3}{6} \ArrowLine(269,-6)(273,-6)
\Text(209,56)[l]{$P_n$} \Text(222,-18)[l]{$q_{n+1}~~c_{n+1}$}

\GCirc(277,-6){4}{0} \LongArrow(277,-2)(277,48)
\Line(277,-56)(277,-10) \Text(265,56)[l]{$\tilde B$}
\Text(273,-66)[l]{$B$}

\end{picture}
\end{center}
\caption{Schematic representation of the process $A+B\rightarrow
\tilde A+P_1+\dots+\tilde B$ in the MRK. The zig-zag lines
represent Reggeized gluon exchange; the black circles denote the
Reggeon vertices; $q_i$ are the Reggeon momenta and $c_i$ are the
colour indices.}
\end{figure}

Denoting momenta of the final particles $k_{i}$, $i=0\div n+1$,
\begin{equation}
k_{i}=\beta _{i}p_{1}+\alpha _{i}p_{2}+k_{i\perp
}~,\;\; s\alpha _{i}\beta _{i%
}=k_{i}^{2}-k_{i\perp }^{2}=k_{i}^{2}+\vec{k}_{i%
}^{~2}~, \label{sud_k_i}
\end{equation}
we have in the MRK
\begin{equation}
\alpha _{0} \ll \alpha _{1}\dots \ll \alpha _{n}\ll \alpha
_{n+1}~, \;\;\; \beta_{n+1} \ll \beta _{n}\dots \ll \beta _{1}\ll
\beta _{0}~. \label{mrk 1}
\end{equation}
Eqs.~(\ref{sud_k_i}) and (\ref{mrk 1}) ensure that  the squared
invariant masses
\begin{equation}
s_{i}=(k_{i-1}+k_{i})^{2}\approx s\beta_{i-1}\alpha _{i}=
\frac{\beta_{i-1}}{\beta_{i}}(k_{i}^{2}+\vec{k}_{i}^{~2})
\label{s_i}
\end{equation}
are large compared with the squared transverse momenta, which are
supposed to be limited (not growing with $s$):
\begin{equation}
s_{i}\gg \vec{k}_{i}^{2}\sim \mid t_{i}\mid =\mid q_{%
i}^{2}\mid ~, \label{mrk 2}
\end{equation}
where
\begin{equation}
t_{i}=q_{i}^{2}\approx q_{i\perp }^{2}=-\vec{q}_{i}^{~2}~,
\label{z9}
\end{equation}
and the product of $s_{i}$ is proportional to $s$:
\begin{equation}
\prod\limits_{i=1}^{n+1}s_{i}=s\prod\limits_{i=1}^{n}(k_{i}^{~2}+
\vec{k}_{i}^{2})~. \label{product s i}
\end{equation}

The real part of the production amplitude can be written as (see
\cite{FF98} and references therein)
\begin{equation}
{\cal A}_{AB}^{\tilde{A}\tilde{B}+n} = 2s\Gamma
_{\tilde{A}A}^{c_{1}} \left[\prod_{i=1}^{n}\frac{1}{t_{i}}
\gamma_{c_{i}c_{i+1}}^{P_{i}}(q_{i},q_{i+1})\left( \frac{s_{i}}
{\sqrt{\vec{k}_{i-1}^{2}\vec{k}_{i}^{2}}}\right)^{\omega
(t_{i})}\right] \frac{1}{t_{n+1}}\left(
\frac{s_{n+1}}{\sqrt{\vec{k}_{n}^{2} \vec{k}_{n+1}^{2}}}\right)
^{\omega (t_{n+1})} \Gamma _{\tilde{B}B}^{c_{n+1}}~, \label{A mrk}
\end{equation}
where $\omega(t)$ is called gluon Regge trajectory (although
actually the trajectory is  $ j(t) = 1 + \omega(t)); \;\;
\Gamma^{c}_{\tilde A A}$ are the particle-particle-Reggeon (PPR)
vertices (they are called also scattering vertices),  i.e.  the
effective vertices for $A \rightarrow \tilde A$ transition due to
interaction with  Reggeized gluon; $c$ is the colour index of this
gluon; $\; \gamma _{c_{i}c_{i+1}}^{P_{i}}(q_{i},q_{i+1})$ are the
Reggeon-Reggeon-particle (RRP) vertices (we'll call them also
production vertices), i.e. the effective vertices for production
of particles $P_i$ with momenta $k_{i}$=$q_{i}-q_{i+1}$ in
collisions of Reggeons with momenta $q_{i}$ and $-q_{i+1}$ and
colour indices $c_{i}$ and $c_{i+1}$. It is clear that in the MRK
only gluons can be produced, so that  all $P_i$ must be gluons.
All scattering vertices $\;\Gamma^c_{P'P},\;$, the gluon
production vertex $\;\gamma^G_{ij}\;$ and the gluon Regge
trajectory  are known now in the NLO (see,
e.g.~\cite{FF98},\cite{FP02} for references), as it is required in
the NLA (in this approximation, in (\ref{A mrk}) either one of the
vertices or one of the trajectory functions must be taken in the NLO).

Note that in the amplitude ${\cal A}_{AB}^{\tilde{A}\tilde{B}+n}$
there are contributions of various colour states and signatures in
the $t_i$-channels, so that, strictly speaking, on the L.H.S. of
(\ref{A mrk}) we should indicate that only  contributions of a
colour octet with negative signature are retained. But since in
this paper we are interested only in such contributions, here and
below we have omitted this indication, in order not to introduce unnecessary
complications. Therefore, in the following it will be understood
that the $t$-channel exchanges belong to colour octet and negative
signature (i.e. the gluon quantum numbers). We remind the reader
that in each order of perturbation theory
amplitudes with the negative signature do dominate, owing to the
cancellation of the leading logarithmic terms in amplitudes with
the positive signature, which become pure imaginary in the LLA due
to this cancellation.

In the NLA the multi-Regge  form for production amplitudes is
assumed to be valid also in the QMRK. Of course, in order to
escape uncertainties, it is necessary to unambiguously separate
two kinds of kinematics. We adopt the separation used
in~\cite{FF98}: by definition, in the MRK all squared invariant
masses of produced particles are larger than some subsidiary
parameter $s_\Lambda$, which is taken to be sufficiently large,
$s_\Lambda \gg |q^2_\perp|$, where $|q_\perp|$ is a typical value
of transverse momenta. Actually the QMRK can be incorporated into
the MRK if we introduce the notion of a `jet'. By definition, we
call a `jet' either a single particle, or a system of two
particles with its invariant mass being less than $s_\Lambda$.
With this definition, instead of speaking of `production of
particles in the QMRK', we will speak about `production of jets in
the MRK', where one jet contains two particles. The QMRK
amplitudes then have the same form (\ref{A mrk}) as in the MRK,
with one of the vertices $\gamma_{c_{i}c_{i+1}}^{P_{i}}$ or
$\Gamma _{\tilde{P}P}^{c}$ being substituted by a vertex for the
production of a pair of particles. Note that because any
two-particle jet in the unitarity relations leads to loss of a
large logarithm, scales of energies in (\ref{A mrk}) are
unimportant in the NLA for the QMRK amplitudes; moreover, the
trajectory and the vertices are needed there only in LO accuracy.
All  vertices for the production of two-particle jets are  known
in this order (for references see, e.g.~\cite{FKR}).

In the following sections we will discuss energy discontinuities
of multiparticle amplitudes. The multi-Regge form (\ref{A mrk})
does not show the full analytic structure, i.e. the dependence on
the energy $s$ and the different subenergies $s_i$. In order to
explain the analytic content of (\ref{A mrk}), it may be useful to
remind the reader of the general structure of inelastic production
amplitudes in the multi-Regge limit ~\cite{White,Bart,FL}; for
simplicity we will restrict ourselves to the $2 \to 3$ scattering
process. In this case we have three energy variables $s$, $s_1$,
and $s_2$, which, in the double Regge limit, are constrained by
the condition $s_1 s_2 = s \vec{k}_1^2$. According to
~\cite{White}, the $2 \to 3$ amplitude in the double Regge limit
can be written in the factorized form (\ref{A mrk}). However, the
production vertex, in general, has a nontrivial phase structure
and an analytic dependence upon the ratio ${s_1 s_2}/{s} =
\vec{k}_1^2$:
\begin{equation}
{\cal A} = \Gamma(t_1)\;\; |z_1|^{j_1}\;\;
      V^{\tau_1 \tau_2}(t_1,t_2,\eta)\;\; |z_2|^{j_2} \;\;
      \Gamma(t_2)~,
\end{equation}
where, for simplicity, we have suppressed the labels $A, A'$ etc.;
$z_1=\cos \theta_1 \approx {2s_1}/({t_1 - t_2})$ and $z_2=\cos
\theta_2 \approx {2s_2}/({t_2 - t_1})$ denote the cosinus of the
cross channel scattering angles, $\eta\approx {s_1 s_2}/s=
\vec{k}_1^2$ has its origin in the Toller angle, $\tau_1, \tau_2$
are the signatures in the $t_1$ and $t_2$ channels, respectively,
and $j_i=1+\omega_i$. We are interested in gluon quantum numbers,
i.e. $\tau_1=\tau_2=-1$. The production vertex $V$ has the general
structure
\begin{eqnarray}
V^{\tau_1 \tau_2}(t_1,t_2,\eta)= |\eta|^{j_1}
\left(e^{-i\pi j_1} +\tau_1 \right)
\left(e^{-i\pi (j_2 -j_1)} +\tau_1 \tau_2 \right)\;\;
\frac{1}{\omega_1-\omega_2}\;\; \frac{1}{\omega_1}\;\; V_1^{\tau_1 \tau_2}
(t_1,t_2,\eta)\;\; \nonumber\\
+ \;\;|\eta|^{j_2} \left(e^{-i\pi j_2} +\tau_2 \right)
\left(e^{-i\pi (j_1 -j_2)} +\tau_1 \tau_2 \right)\;\;
\frac{1}{\omega_2-\omega_1}\;\; \frac{1}{\omega_2}\;\; V_2^{\tau_1
\tau_2} (t_1,t_2,\eta)~.\hspace{0.2cm}
\end{eqnarray}
It is easy to see, by expansion in powers of $g$, that for
$\tau_1=\tau_2=-1$ the amplitude is real-valued, i.e. the phase
factors can be approximated by $\pm 2$, and by a suitable change
of the energy scales and a redefinition of the vertex factors
$\Gamma$ and $V_1$, $V_2$, one arrives at the form (\ref{A mrk}).
However, at this stage the information on the analytic structure
has been lost. Instead, if we rewrite Eq.~(2.10), with a
redefinition of the vertex factors, as
\begin{eqnarray}
{\cal A}= s^{j_1} s_2^{j_2 -j_1} \left(e^{-i\pi j_1} +\tau_1 \right)
\left(e^{-i\pi (j_2 -j_1)} +\tau_1 \tau_2 \right)\;\;\Gamma(t_1)\;\;
\frac{1}{\omega_1-\omega_2}\;\; \frac{1}{\omega_1}\;\; V_1^{\tau_1 \tau_2}
(t_1,t_2,\eta)\;\; \Gamma(t_2)\nonumber\\
+\;\; s^{j_2} s_1^{j_1-j_2} \left(e^{-i\pi j_2} +\tau_2 \right)
\left(e^{-i\pi (j_1 -j_2)} +\tau_1 \tau_2
\right)\;\;\Gamma(t_1)\;\; \frac{1}{\omega_2-\omega_1}\;\;
\frac{1}{\omega_2}\;\; V_2^{\tau_1 \tau_2} (t_1,t_2,\eta) \;\;
\Gamma(t_2)~, \label{bart} \hspace{0.4cm}
\end{eqnarray}
we can associate the phase factors with the energy factors, and
the dependence on the energy variables becomes transparent: the
first term on the R.H.S. is a function of $s$ and $s_2$ and has
the usual right and left hand cut structure, the second one
depends upon $s$ and $s_1$. Consequently, the single discontinuity
in $s_1$, $disc_{s_1}{\cal A}$, is obtained from the second term
only and within the NLO can be written as\be disc_{s_1} {\cal A}
=2 \pi i s^{j_2} s_1^{j_1-j_2} \left(e^{-i\pi \j_2} + \tau_2
\right)\;\; \Gamma(t_1)\;\; \frac{1}{\omega_2}\;\; V_2^{\tau_1
\tau_2} (t_1,t_2,\eta) \;\; \Gamma(t_2)~, \ee whereas the
discontinuity in $s$, $disc_s {\cal A}$,  has contributions from
both terms (this structure has been used to calculate separately
the two vertex functions $V_1$ and $V_2$  in the LO in~\cite{Bart}
and in the NLO in~\cite{FL}). Moreover, the form (2.12) also
allows to compute double discontinuities, e.g. $disc_s\;disc_{s_1}
{\cal A}$. The most essential property of this representation is
the absence of discontinuities in overlapping channels (i.e.
$disc_{s_1}\;disc_{s_2} {\cal A}\;=\;0$) and in the decoupling of
singularities in $s$ and $s_1$ or $s$ and $s_2$. Now it is easy to
show that  at $\tau_1=\tau_2=-1$ the sum of two single
discontinuities, $disc_{s_1} + disc_s$, in the approximation where
the phase factors are taken to be equal to $\pm 2$, is
proportional to the full amplitude again:
\begin{equation}
\left(disc_s + disc_{s_1}\right) {\cal A} = - \omega_1 \pi i \;\;{\cal A},
\end{equation}
and we do not need to discuss the two pieces of the production
vertex, $V_1$ and $V_2$, separately. An analogous discussion holds
for the $2 \to 4$ production amplitude ~\cite{Bart}: the $2 \to 4$
amplitude can be written as a sum of 5 terms, each of which has a
simple analytic structure in a certain subset of energy variables.
Single energy discontinuities pick out only a few of these terms,
whereas certain sums of single discontinuities can be shown to be
proportional to the full amplitude (see below).

In the following we will argue that the simple relation (2.14)
(and its generalizations to the two gluon production) can be
obtained directly from the inspection of QCD perturbation theory,
and we then will use these equations for the derivation of
bootstrap conditions.

\section{Bootstrap relations}

Let us first recall the derivation of the bootstrap relation for the
elastic amplitude ${\cal A}_{AB}^{A'B'}$. In the limit of large
$s$ the radiative corrections of order $\alpha_S^k$ to this
amplitude divided by $s$ can depend on $s$ only in the form
$\left(\ln^n(-s)+\ln^n s\right)$ with $n\leq k$ (remember that we
consider negative signature). With the NLO accuracy we can put
\be
\frac{1}{-2\pi i}disc_s \;(\ln^n(-s)+\ln^n
s)=\frac{1}{2}\frac{\partial }{\partial  \ln s}\Re
\left[\ln^n(-s)+\ln^n s\right]~, \label{dis1}
\ee
where $disc_s$
denotes the $s$-channel discontinuity, and $\Re$ indicates the real part.
Therefore in the NLA  we have
\be \frac{1}{-2\pi i s} disc_{s} \;
{\cal A}_{AB}^{A'B'}=\frac{1}{2}\frac{\partial}{\partial \ln s}\Re
\; \left[\frac{1}{s}{\cal A}_{AB}^{A'B'}\right]~.\label{dis2}
\ee
Substituting the Reggeized form (\ref{A mrk}) into
$\Re \,{\cal
A}_{AB}^{A'B'}$ we obtain the bootstrap relation:
\be
\frac{1}{-2\pi i s} disc_{s} {\cal
A}_{AB}^{A'B'}=\frac{1}{2}\omega(t)\Re \; \left[\frac{1}{s}{\cal
A}_{AB}^{A'B'}\right]~.\label{boot el}
\ee
The important point is
that the $s$-channel discontinuity on the L.H.S. of the relation
(\ref{boot el}) can be calculated by inserting the amplitude of the
form (\ref{A mrk}) into the unitarity condition. Since the
amplitudes are expressed through the vertices of the Reggeon
interactions and the gluon Regge trajectory, the relation
(\ref{boot el}) imposes a series of restrictions on the vertices
and trajectory, which have been formulated as bootstrap conditions
for the color octet impact factors and for the BFKL
kernel~\cite{FF98}. Note that in order to obtain these conditions in the
NLO it is sufficient to retain, on both sides of the relation
(\ref{boot el}), only terms linear in $\ln s$.

The bootstrap relations for the one-gluon production amplitude $A
+ B \; \rightarrow \;  A' + G + B'$ in the MRK are derived in a
similar way, although this case is slightly more complicated. As
discussed before, in this amplitude there are three energy
variables, $s_1=(p_{A'}+k)^2,\; s_2=(k+p_{B'})^2,\;$ ($k$ is the
momentum of the produced gluon), and $s=(p_A+p_B)^2\simeq
(p_{A'}+p_{B'})^2$, and two momentum transfers,
$t_1=(p_A-p_{A'})^2\;$ and $t_2=(p_B-p_{B'})^2.\;$ Recall that we
are considering only negative signatures in both $t$- channels,
i.e. the part of the amplitude which is antisymmetric with respect
to any of the substitutions $s_1\rightarrow -s_1$ and
$s_2\rightarrow -s_2$. Due to the relation ${s_1s_2}=s\vec
k^{~2}$,  which is fulfilled in all physical channels, this part
is also antisymmetric with respect to $s\rightarrow -s$. In the
MRK logarithms of all energy  variables are considered to be large
(i.e. on the same footing as $\ln s$ in the elastic amplitude).
Let us consider the discontinuities of the amplitude in these
variables (for brevity we sometimes call all these discontinuities
$s$-channel ones). In general the determination of discontinuities
of inelastic amplitudes is not simple. In particular, when
calculating the discontinuity in one of the energy variables one
needs to care on which edges of their cuts are the other
variables. It is related to the existence of double
discontinuities: e.g., the discontinuity in $s$ can have, in turn,
discontinuities in $s_1$ or $s_2$, so that the single
discontinuities are not pure imaginary. Fortunately, for our
purposes it is sufficient to consider only imaginary parts of the
discontinuities, so that this complication is irrelevant. The
second complication is that, because of the relation
${s_1s_2}=s\vec k^{~2}$, the large logarithms compensate each
other if they enter in the combination $\ln s_1+\ln s_2-\ln s =\ln
\vec k^{~2}$, so that radiative corrections of some fixed order in
$\alpha_S$ can contain each of the large logarithms in any power
through the dependence on $\vec k^{~2}$. Therefore equalities like
(\ref{dis2}) are evidently absent. This difficulty can be overcome
noticing that there are combinations of the discontinuities in
which contributions related to dependence on $\vec k^{~2}$ cancel.
Indeed, the sums $(disc_{s_{1,2}}+disc_{s}) F(s_1s_2/s)$ are zero
(as well as any superposition of these sums). Note that just these
sums, as already indicated in (2.14), are proportion within the
NLA to the full amplitude. The most general form for the
dependence of radiative corrections (to the amplitude divided by
$s$) on the energy variables is a superposition of functions of
$\vec k^{~2}$ multiplied by powers of large logarithms.  Then, by
noticing that a discontinuity of a product of two functions is
expressed through the discontinuities of these functions as
\begin{equation}
f_+g_+-f_-g_-
=\frac{1}{2}(f_+-f_-)(g_++g_-)+\frac{1}{2}(f_++f_-)(g_+-g_-)~,
\label{discfg}
\end{equation}
 we conclude that calculating
$(disc_{s_{1,2}}+disc_{s}){\cal A}_{AB}^{A'GB'}$ we can ignore
the analytical properties of the functions of $\vec k^{~2}$  and
take their real parts.

Now, if the variables   $s_1$,  $s_2$ and  $s$ do not enter into
the combination $s_1s_2/s=\vec k^{~2}$, they can appear in the
radiative corrections of order $\alpha_S^k$ to the amplitude
divided by $s$ only as $\hat{\cal
S}\ln^{n_1}(-s_1)\ln^{n_2}(-s_2)\ln^{n_3}(\pm s)$, with
$n_1+n_2+n_3=n\leq k$, where $\hat{\cal S}$ is the operator of
symmetrization with respect to exchanges $ s_1\leftrightarrow
-s_1~, \;\;s\leftrightarrow -s$ and $ s_2\leftrightarrow -s_2~,
\;\;s\leftrightarrow -s~.$  Note that the terms containing
products of $\ln(-s_i)\ln(s_i)$, where $s_i$ can be $s_1$,  $s_2$
or  $s$, are {forbidden}, on the same ground as the terms
containing $\ln(-s)\ln(s)$ are {forbidden} in the elastic
amplitudes. In analogy to our treatment of the elastic case, in
our actual calculation of energy discontinuities in the next
section  we will restrict ourselves to terms which, in the
discontinuities, are linear in large logarithms, i.e. we need only
$n\leq 2$. But the bootstrap relations can easily be derived
without this restriction, as it is done below. Since in the NLO we
need to {keep only the first two leading total powers} of n,
calculating the {imaginary} part of the discontinuity in any of
the variables $s_1$, $s_2$ or  $s$ we can take only real parts of
logarithms of the other variables. It means that with our accuracy
\[ \Re \; \left[\frac{1}{-2\pi i}
disc_{s_i} \left(\hat{\cal
S}\ln^{n_1}(-s_1)\ln^{n_2}(-s_2)\ln^{n_3}(\pm s)\right)\right]
\]
\be =\frac{1}{2}\frac{\partial}{\partial \ln s_i}\Re
\;\left[\hat{\cal S}\ln^{n_1}(-s_1)\ln^{n_2}(-s_2)\ln^{n_3}(\pm
s)\right]~,
\ee where $s_i$ can be  $s_1$,  $s_2$ or  $s$, and  the
partial derivative is taken at fixed $s_j\neq s_i$.

Therefore we have, for example, \be \Re \; \left[\frac{1}{-2\pi i
s} \left(disc_{s_1}+disc_{s}\right) {\cal
A}_{AB}^{A'GB'}\right]=\frac{1}{2} \left(\frac{\partial}{\partial
\ln s_1}+\frac{\partial}{\partial \ln s}\right) \Re \;
\left[\frac{1}{s}{\cal A}_{AB}^{A'GB'}(s_1,s_2,s)\right]~,
\label{sum discs=sum der} \ee where on the R.H.S. the first
derivative is taken at fixed $s_2$ and $s$, and the second one at
fixed  $s_1$ and  $s_2$. Using the equality \be
\left(\frac{\partial}{\partial \ln s_1}+\frac{\partial}{\partial
\ln s}\right)f(s_1,s_2, s)= \frac{\partial}{\partial \ln
s_1}f(s_1,s_2, \frac{s_1 s_2}{\vec k^{~2}})~, \ee we arrive at \be
\Re \; \left[\frac{1}{-2\pi i s}\left(disc_{s_1}+disc_{s}\right)
{\cal A}_{AB}^{A'GB'}\right]=\frac{1}{2} \frac{\partial}{\partial
\ln s_1} \Re\;\left[\frac{1}{s}{\cal A}_{AB}^{A'GB'}\right]~, \ee
where on the R.H.S. the amplitude is considered as a function of
$s_1,\;\;s_2$ and $\vec k^{~2}$~.

The {requirement of the Reggeized form} (\ref{A mrk}) of the
amplitude on the R.H.S.  gives us the {bootstrap relation}: \be
\Re\;\left[ \frac{1}{-2\pi i}\left(disc_{s_1}+disc_{s}\right)
{\cal A}_{AB}^{A'GB'}\right]=\frac{1}{2} \omega(t_1)\Re\;{\cal
A}_{AB}^{A'GB'}~, \label{s1+s} \ee which coincides with (2.14). In
the same way we obtain \be \Re\;\left[\frac{1}{-2\pi i}
\left(disc_{s_2}+disc_{s}\right) {\cal
A}_{AB}^{A'GB'}\right]=\frac{1}{2} \omega(t_2)\Re\;{\cal
A}_{AB}^{A'GB'}~. \label{s2+s} \ee

It is known \cite{FF98} that the bootstrap relation (\ref{boot
el}) for elastic scattering in the NLO leads to the bootstrap
conditions for the impact factors and kernel.  As we will see in
Section 5, relations (\ref{s2+s}) and (\ref{s1+s}) (actually they
are equivalent, so that the consideration of only one of them is
sufficient) require the strong form of these conditions. Besides
this, these relations give a completely  new condition, which
involves new "impact factors". This new condition appears in a
"weak" form, analogous to the form of the  conditions for the
impact factors and kernel obtained from the elastic bootstrap. So
the bootstrap relations for one-gluon production play a two-fold
role: they strengthen the conditions imposed by the elastic
bootstrap, and they give a new one. One might expect that the
history will repeat itself with the addition of each next gluon in
the final state. If this would be so, we had to consider bootstrap
relations for production of arbitrary number of gluons, and we
would obtain an infinite number of bootstrap conditions.
Fortunately, history is repeated only partly: it turns out that
already the bootstrap relations for two-gluon production only
strengthen the form of the new condition imposed by the bootstrap
for one-gluon production, and so the production of two gluons does
not require new conditions. Therefore it is sufficient to consider
the bootstrap relations for amplitudes of two-gluon production.
They are derived in a way similar to the one-gluon case, although
there are complications related to the larger number of energy
variables and the larger number of combinations of energy
variables which do not grow with $s$.  Applying the notations of
Section 2 to the production of two gluons, i.e. $n=2$, we have six
energy variables $s_{i,j}=(k_i+k_j)^2~, \;\;j>i=0\div 2\;\;
(s_{0,3}=s),$ and two squared transverse momenta of produced
gluons $\vec k^{~2}_1, \vec k^{~2}_2$ with four relations between
them:
\begin{equation}
\frac{s_{0,1}s_{1,2}}{s_{0,2}}=\frac{s_{0,1}s_{1,3}}{s_{0,3}}=\vec
k^{~2}_1, \;\;\;
\frac{s_{1,2}s_{2,3}}{s_{1,3}}=\frac{s_{0,2}s_{2,3}}{s_{0,3}}
=\vec k^{~2}_2~ \label{sisjsij}
\end{equation}
(actually only three of these relations are independent, since
ratios of the first two terms in the first equality is identically
equal to analogous ratios for the second equality).  Similar to the
production of one gluon, in order to avoid
the need of going beyond the logarithmic dependence of the
production amplitude on the energy variables we have to take
definite combinations of discontinuities in these variables. The
combinations must satisfy the requirement that for functions of
$\vec k^{~2}_1, \vec k^{~2}_2$ they give zero. Let us take, for
example, the sum of discontinuities in the channels
${s_{2,3}},{s_{1,3}}$ and ${s}$. From (\ref{sisjsij}) it is
readily seen that
\begin{equation}
(disc_{s_{2,3}}+disc_{s_{1,3}}+disc_{s}) F(\vec k^{~2}_1, \vec
k^{~2}_2) =0 \label{discF(k1k2)}~.
\end{equation}
Using this and (\ref{discfg}) we conclude that calculating the sum
of the amplitude discontinuities in the channels $s_{2,3},
s_{1,3}$ and ${s}$ we can ignore the analytical properties of the
amplitude in the variables $\vec k_1^{~2}$ and $\vec k_1^{~2}$.
Then, quite analogous to (\ref{sum discs=sum der}), we obtain
\[
\Re \; \left[\frac{1}{-2\pi i s}
\left(disc_{s_{2,3}}+disc_{s_{1,3}}+disc_{s}\right) {\cal
A}_{AB}^{A'G_1G_2B'}\right]
\]
\begin{equation}
=\frac{1}{2} \left(\frac{\partial}{\partial \ln
s_{2,3}}+\frac{\partial}{\partial \ln
s_{1,3}}+\frac{\partial}{\partial \ln s}\right) \Re \;
\left[\frac{1}{s}{\cal A}_{AB}^{A'G_1G_2B'}\right]~, \label{sum
discs=sum der for two gluons}
\end{equation}
where the amplitude on the R.H.S. is considered as a function of
${s_{2,3}},{s_{1,3}}, s$ and $\vec k^{~2}_1, \vec k^{~2}_2$.
Passing on to ${s_{2,3}}\equiv s_3~,\;\;{s_{1,2}}\equiv s_3~,\;\;
s_{0,1}\equiv s_3~\;\;$ and $\vec k^{~2}_1~, \;\;\vec k^{~2}_2$ as
independent variables, and using the requirement of the Reggeized
form  (\ref{A mrk}) we arrive at  (cf.~(\ref{s2+s})):
\begin{equation}
\Re\;\left[ \frac{1}{-2\pi
i}\left(disc_{s_{2,3}}+disc_{s_{1,3}}+disc_{s}\right) {\cal
A}_{AB}^{A'G_1G_2B'}\right]=\frac{1}{2} \omega(t_3)\Re\;{\cal
A}_{AB}^{A'G_1G_2B'}~. \label{s23+s13+s}
\end{equation}
It is one of three independent {bootstrap relations} for the
two-gluon production amplitude. The other two  relations,
\[
\Re\;\left[\frac{1}{-2\pi
i}\left(disc_{s_{0,1}}+disc_{s_{0,2}}+disc_{s}\right) {\cal
A}_{AB}^{A'G_1G_2B'}\right]=\frac{1}{2} \omega(t_1)\Re\;{\cal
A}_{AB}^{A'G_1G_2B'}~,
\]
\begin{equation}
\Re\;\left[ \frac{1}{-2\pi
i}\left(disc_{s_{1,2}}+disc_{s_{1,3}}-disc_{s_{0,1}}\right) {\cal
A}_{AB}^{A'G_1G_2B'}\right]=\frac{1}{2}
(\omega(t_2)-\omega(t_1))\Re\;{\cal A}_{AB}^{A'G_1G_2B'}~,
\label{s01+s02+s}
\end{equation}
are obtained in the same way. We finally note that the relations
(\ref{s23+s13+s}) and (\ref{s01+s02+s}) can be derived easily also
from a representation analogous to (\ref{bart}).

\section{Calculation of $s$-channel discontinuities}

\subsection{Discontinuity of elastic amplitudes}

It is worth-while to start with the elastic amplitude ${\cal
A}_{AB}^{A'B'}$, at least in order to introduce the notions of impact
factors and of the BFKL kernel (of course, in the colour octet channel).
After this the calculation of the discontinuity of the elastic
amplitude can be generalized to inelastic amplitudes in a relatively
simple  way.

The $s$-channel discontinuity is calculated with the help of the
unitarity relation, using on the R.H.S. of this relation the
Reggeized form (\ref{A mrk}) of the amplitudes.  In Eq.~(\ref{A
mrk}) only the real part of the amplitude is given, and we have
omitted the symbol of the real part. In order to simplify the
notations, in the following we shall omit this symbol  without
further notice. Fortunately, within the NLO accuracy only real
parts are important for the calculation of the discontinuity.  We
will content ourselves with terms of the zeroth and first power of
$\ln s$ in the discontinuity, since, as we shall see, this is
sufficient for the derivation of the bootstrap conditions. In the
NLA these terms can come from intermediate states with two, three
and four jets (see Fig.~2). Indeed, in the LLA each additional
gluon in the intermediate state gives a large logarithm, therefore
for $n$ jets in the intermediate state there are at least $n-2$
logarithms. Since we work in the NLA, in each order of
perturbation theory in $g^2$ we have to retain only the leading
and next-to leading terms. So it is clear that the five-jet
contribution is irrelevant for us.

\begin{figure}[htb]
\begin{center}
\begin{picture}(400,400)(-40,-380)

\CBoxc(30,0)(0.8,50){0}{0}
\Line(24,-19)(30,-13)
\Line(36,-19)(30,-13)
\Line(24,7)(30,13)
\Line(36,7)(30,13)
\LongArrow(30,33)(30,52)
\ArrowLine(30,-52)(30,-33)
\ZigZag(34,-29)(82,-29){3}{5}
\ZigZag(34,29)(82,29){3}{5}
\ArrowLine(82,-29)(86,-29)
\ArrowLine(82,29)(86,29)
\GCirc(30,-29){4}{0}
\GCirc(30,29){4}{0}
\Text(24,62)[l]{$A^{\prime}$}
\Text(24,-62)[l]{$A$}
\Text(12,-10)[l]{${\tilde{A}}$}
\Text(52,40)[l]{$r^{\prime}~c^{\prime}$}
\Text(52,-40)[l]{$r~c$}
\Text(57,-104)[l]{$a$}

\DashLine(15,0)(105,0){4}

\GCirc(90,-29){4}{0}
\GCirc(90,29){4}{0}
\CBoxc(90,0)(0.8,50){0}{0}
\Line(84,-19)(90,-13)
\Line(96,-19)(90,-13)
\Line(84,7)(90,13)
\Line(96,7)(90,13)
\LongArrow(90,33)(90,52)
\ArrowLine(90,-52)(90,-33)
\Text(84,62)[l]{$B^{\prime}$}
\Text(84,-62)[l]{$B$}
\Text(108,-10)[r]{${\tilde{B}}$}

\CBoxc(160,0)(0.8,50){0}{0} \Line(154,-19)(160,-13)
\Line(166,-19)(160,-13) \Line(154,7)(160,13) \Line(166,7)(160,13)
\LongArrow(160,33)(160,52) \ArrowLine(160,-52)(160,-33)
\ZigZag(164,-29)(212,-29){3}{5} \ZigZag(164,29)(212,29){3}{5}
\ArrowLine(212,-29)(216,-29) \ArrowLine(212,29)(216,29)
\GCirc(160,-29){4}{0} \GCirc(160,29){4}{0}
\Text(154,62)[l]{$A^{\prime}$} \Text(154,-62)[l]{$A$}
\Text(142,-10)[l]{${\tilde{A}}$}
\Text(182,40)[l]{$r_1^{\prime}~c_1^{\prime}$}
\Text(182,-40)[l]{$r_1~c_1$}

\GCirc(220,29){4}{0}
\GCirc(220,-29){4}{0}
\CBoxc(220,0)(0.8,50){0}{0}
\Line(214,-19)(220,-13)
\Line(226,-19)(220,-13)
\Line(214,7)(220,13)
\Line(226,7)(220,13)
\Text(225,-10)[l]{$J$}

\Text(218,-114)[l]{$b$}

\DashLine(145,0)(295,0){4}

\ZigZag(224,-29)(272,-29){3}{5} \ZigZag(224,29)(272,29){3}{5}
\ArrowLine(272,-29)(276,-29) \ArrowLine(272,29)(276,29)
\GCirc(280,-29){4}{0} \GCirc(280,29){4}{0}
\CBoxc(280,0)(0.8,50){0}{0} \Line(274,-19)(280,-13)
\Line(286,-19)(280,-13) \Line(274,7)(280,13) \Line(286,7)(280,13)
\LongArrow(280,33)(280,52) \ArrowLine(280,-52)(280,-33)
\Text(276,62)[l]{$B^{\prime}$} \Text(275,-62)[l]{$B$}
\Text(298,-10)[r]{${\tilde{B}}$}
\Text(246,40)[l]{$r_2^{\prime}~c_2^{\prime}$}
\Text(247,-40)[l]{$r_2~c_2$}

\CBoxc(65,-240)(0.8,50){0}{0} \Line(59,-259)(65,-253)
\Line(71,-259)(65,-253) \Line(59,-233)(65,-227)
\Line(71,-233)(65,-227) \LongArrow(65,-207)(65,-189)
\ArrowLine(65,-291)(65,-273) \ZigZag(69,-211)(117,-211){3}{5}
\ZigZag(69,-269)(117,-269){3}{5} \ArrowLine(117,-211)(121,-211)
\ArrowLine(117,-269)(121,-269) \GCirc(65,-269){4}{0}
\GCirc(65,-211){4}{0} \Text(61,-178)[l]{$A^{\prime}$}
\Text(59,-302)[l]{$A$} \Text(47,-250)[l]{${\tilde{A}}$}
\Text(83,-200)[l]{$r_1^{\prime}~~c_1^{\prime}$}
\Text(83,-280)[l]{$r_1~~c_1$}

\CBoxc(125,-240)(0.8,50){0}{0} \Line(119,-259)(125,-253)
\Line(131,-259)(125,-253) \Line(119,-233)(125,-227)
\Line(131,-233)(125,-227) \GCirc(125,-269){4}{0}
\GCirc(125,-211){4}{0} \ZigZag(129,-211)(177,-211){3}{5}
\ZigZag(129,-269)(177,-269){3}{5} \ArrowLine(177,-211)(181,-211)
\ArrowLine(177,-269)(181,-269)
\Text(143,-200)[l]{$r_2^{\prime}~~c_2^{\prime}$}
\Text(143,-280)[l]{$r_2~~c_2$} \Text(130,-250)[l]{$J_1$}

\CBoxc(185,-240)(0.8,50){0}{0}
\Line(179,-259)(185,-253)
\Line(191,-259)(185,-253)
\Line(179,-233)(185,-227)
\Line(191,-233)(185,-227)
\GCirc(185,-269){4}{0}
\GCirc(185,-211){4}{0}
\ZigZag(189,-211)(237,-211){3}{5}
\ZigZag(189,-269)(237,-269){3}{5}
\ArrowLine(237,-211)(241,-211)
\ArrowLine(237,-269)(241,-269)
\Text(190,-250)[l]{$J_2$}

\Text(156,-354)[l]{$c$}

\DashLine(45,-240)(265,-240){4}

\GCirc(245,-269){4}{0} \GCirc(245,-211){4}{0}
\CBoxc(245,-240)(0.8,50){0}{0} \Line(239,-259)(245,-253)
\Line(251,-259)(245,-253) \Line(239,-233)(245,-227)
\Line(251,-233)(245,-227) \LongArrow(245,-207)(245,-189)
\ArrowLine(245,-291)(245,-273) \Text(241,-178)[l]{$B^{\prime}$}
\Text(239,-302)[l]{$B$} \Text(263,-250)[r]{${\tilde{B}}$}
\Text(203,-200)[l]{$r_3^{\prime}~~c_3^{\prime}$}
\Text(203,-280)[l]{$r_3~~c_3$}

\end{picture}
\end{center}
\caption[]{Schematic representation of the contributions to the
$s$-channel discontinuity of the elastic amplitude ${\cal
A}_{AB}^{A'B'}$: a) two-jet, b) three-jet and c) four-jet.}
\end{figure}
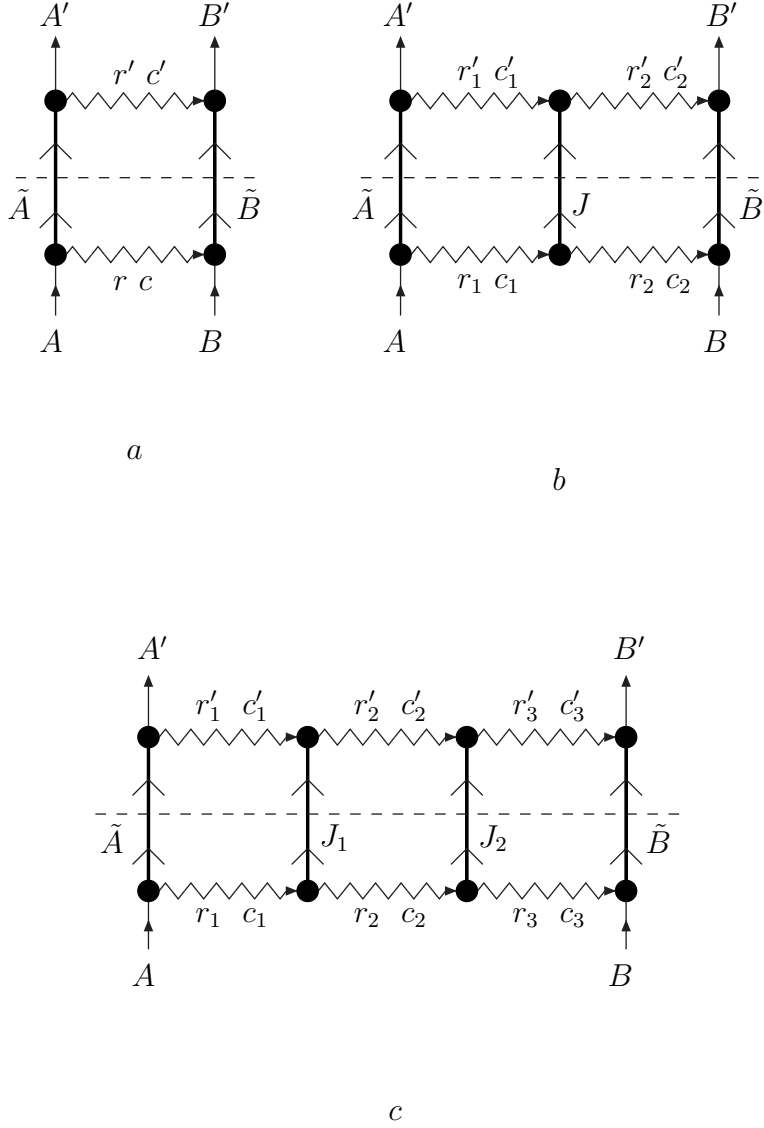

The  unitarity relation with two jets in the intermediate state
gives  (see Fig.~2a))
\begin{equation}
\frac{1}{-2\pi i}disc^{(2\Lambda)}_{s}{\cal A}_{AB}^{A'B'}=
-\frac{1}{2\pi}\sum_{ \tilde A  \tilde B} {\cal A}_{AB}^{ \tilde A
 \tilde B}\; {\cal A}_{ \tilde A  \tilde B}^{A'B'}
 d\rho_{ \tilde A  \tilde B}~, \label{disc 2}
\end{equation}
where  the superscript $\Lambda$ indicates  that the squared
invariant mass of the jet is less than $s_\Lambda;\;$ $ \tilde A $
and $ \tilde B$ are the jets with momenta $p_{\tilde A}$ and
$p_{\tilde B}$, respectively, $d\rho_{\tilde A  \tilde B}$ denotes
their phase space element. Amplitudes on the R.H.S. of
Eq.~(\ref{disc 2}) are of the form (\ref{A mrk}). Remember that it
is our aim to obtain bootstrap conditions by inserting the
discontinuity into the relation (\ref{boot el}). To compare the
left and right hand sides of the relation (\ref{boot el}) we need
to use the same scale of energy on both sides. We choose the scale
$-t=-q_\perp^2$, which is natural for the amplitude ${\cal
A}_{AB}^{A'B'}$, but not for the amplitudes ${\cal A}_{AB}^{
\tilde A \tilde B}$ and ${\cal A}_{ \tilde A \tilde B}^{A'B'}$.
Passing to this scale in Eq.~(\ref{A mrk}), we obtain
\begin{equation}
{\cal A}^{ \tilde A  \tilde B}_{AB} = \frac{2 s}{{r^2}}~
\Gamma^{c}_{ \tilde A A}(q_t)
\left(\frac{s}{q^{2}_{t}}\right)^{\omega(r^2)} \Gamma^{c}_{ \tilde
BB}(q_t)~,\;\; {\cal A}_{\tilde A\tilde B}^{A'B'} = \frac{2
s}{{r'^2}}~ \Gamma^{c}_{A'\tilde
A}(q_t)\left(\frac{s}{q^{2}_t}\right)^{\omega(r'^2)}
\Gamma^{c}_{B'\tilde B}(q_t)~.\label{2A 2jet}
\end{equation}
Here $r=p_A- p_{ \tilde A }$ and $r'=q-r$ are the transferred
momenta. Remind that $q=p_{A}-p_{A'}$ and we can put $q=q_\perp$.
Since in the unitarity relation essential transverse momenta are
limited (i.e. they do not grow with $s$), we can put also
$r=r_\perp~, r'=r'_\perp$. We have introduced the notation $p_t
\equiv \sqrt{-p^2_\perp}$ for any $p$, and the scattering vertices
at the scale $q_t$. For any transition $P\rightarrow \tilde P$
with momentum transfer $p$ these vertices are defined as
\begin{equation}
 \Gamma^{c}_{ \tilde P P}(q_t)
=\Gamma^{c}_{ \tilde P P}
\left(\frac{q_t}{p_{t}}\right)^{\omega(p^2)}~.\;\; \label{Gamma
scaled}
\end{equation}
In order to write formulas in a
simple form we do not perform an explicit expansion in the
coupling $g^2$. In practice, however, the expansion is always assumed,
and only the NLA accuracy is needed. Therefore we have
\[
\Gamma^{c}_{ \tilde P P}(q_t) =\Gamma^{c}_{ \tilde P P}
\left(\frac{q_t}{p_{t}}\right)^{\omega(p^2)}=\Gamma^{c}_{ \tilde P
P}\left[1+ {\omega(p^2)}\ln\left(\frac{q_t}{p_{t}}\right)\right]
\]
\begin{equation}
=\Gamma^{c}_{ \tilde P P}+ \Gamma^{c(B)}_{ \tilde P
P}{\omega(p^2)}\ln \left(\frac{q_t}{p_{t}}\right)~.\label{Gamma
scaled NLA}
\end{equation}
Furthermore, since we are going to retain only the zeroth and the first
power of $\ln s$ we can put
\begin{equation}
\left(\frac{s}{q^{2}_{t}}\right)^{\omega(r^2)}
\left(\frac{s}{q^{2}_t}\right)^{\omega(r'^2)} =1+\Omega
Y~,\label{OmegaY}
\end{equation}
where we have introduced the notations
\be
\Omega=\omega(r^2)+\omega(r'^2)~, \;\;Y=\ln
\left(\frac{s}{q^{2}_{t}}\right) \label{Omega}\ee
used below.

It is convenient to write down the
phase space element $d\phi_{J}$ for a jet $J$ with total momentum
$k_J$ consisting of particles with momenta $l_i$:
\begin{equation}
d\phi_{J}=\frac{d k^2_J}{2\pi}\theta(s_\Lambda -k^2_J)(2\pi)^{D}
\delta^D(k_J-\Sigma_i l_i)\prod_i\frac{d^{D-1}l_{i}}{\left( 2\pi
\right)^{D-1}2\epsilon _{i}}~. \label{rho jet}
\end{equation}
For the phase space element for two produced jets $d\rho_{
\tilde A  \tilde B}$ in Eq.~(\ref{disc 2}) we have
\begin{equation}
d\rho_{ \tilde A \tilde B}=d\phi_{ \tilde A }d\phi_{ \tilde
B}\frac{d^{D-2}r_{\perp}}{2s(2\pi)^{D-2}}~.\label{rho AB}
\end{equation}
In Eq.~(\ref{disc 2}) the sum extends over colours and
polarizations of the intermediate particles. It is performed
independently for each jet. Note that the projection onto the
antisymmetric colour octet state in the $t$-channel is always
understood. Making this projection explicitly, we define the
non-subtracted impact factors as
\[
{\Phi}_{A'A}^{i(\Lambda)}(r_\perp,
r'_\perp)=~i\frac{f^{icc'}}{N_c}\sum_{ \tilde A  }\int
\Gamma^{c}_{ \tilde A  A }(q_t) \Gamma^{c'}_{A' \tilde A  }(q_t)
 d\phi_{ \tilde A
}~,
\]
\begin{equation}
{\Phi}_{B'B}^{i(\Lambda)}(-r_\perp,
-r'_\perp)=~i\frac{f^{icc'}}{N_c}\sum_{ \tilde B  }\int
\Gamma^{c}_{ \tilde B  B }(q_t) \Gamma^{c'}_{B' \tilde B  }(q_t)
 d\phi_{ \tilde B
}~. \label{impact un}
\end{equation}

In order to simplify the representation of our results for the
discontinuities it is convenient to introduce {operators in the
transverse momentum representation}. From the $t$-channel point of
view we have to consider two interacting Reggeized gluons (see
Fig.~2) with "coordinates" ${\vec r}$ and $\vec q-\vec r$ in the
transverse momentum space ($\vec q$ is the total transverse
momentum in the $t$-channel). Let us introduce $\hat{\vec r}$ as
the operator of "coordinate" of one of the Reggeized gluons in the
transverse momentum space: $ \hat{\vec r}\: |\vec q_i\rangle =
\vec q_i|\vec q_i\rangle~. $ The total transverse momentum $\vec
q\;$  is considered as a $c$-number. With the normalization $
\langle\vec q_1|\vec q_2\rangle = \vec q_1^{\:2}(\vec q_1 - \vec
q)^{\:2}\delta^{(D-2)}(\vec q_1 - \vec q_2)~$ we define
\begin{equation} \langle\Psi_2|\Psi_1\rangle =
\int\frac{d^{D-2}r}{\vec r^{~2}(\vec q-\vec r)^2} \langle
\Psi_2|\vec r\rangle\langle\vec r| \Psi_1\rangle~.
\end{equation}
In this formalism the {impact factors} $\Phi_{A^\prime A}$ and
$\Phi_{B^\prime B}$ appear as the ''wave functions'' of the
$t$-channel states $\langle {A^\prime A}|$ and  $|{ B^\prime B
}\rangle~$,  respectively, and the {BFKL kernel} ${\cal K}(\vec
r_2, \vec r_1, \vec q)$ as the ``matrix element'' $ \langle\vec
r_1|\hat{\cal K}|\vec r_2\rangle$. Since the $t$-channel is
assumed to be in a colour octet state, the impact factors carry a
colour index. For simplicity, we have omitted this index, and in
the following we often do the same whenever possible (the same
applies to colour indices of Reggeon vertices). It is worthwhile
to mention that impact factors are assumed to be symmetric under
the exchange of the two gluon momenta: $r_{\perp} \leftrightarrow
r^{\prime}_\perp$. For the quark and gluon impact factors this
property is fulfilled automatically. It is not so in more
complicated cases; therefore, in the general case, symmetrization
with respect to the exchange $r_\perp\leftrightarrow
r^{\prime}_\perp$ on the R.H.S. of Eq.~(\ref{impact un}) is
understood.

With these definitions we can write
\begin{equation}
\frac{1}{-2\pi i}disc^{(2\Lambda)}_{s}{\cal A}_{AB}^{A'B'}
=\frac{2 sN_c}{(2\pi)^{D-1}}~ \langle
A'A^{(\Lambda)}|1+\hat{\Omega} Y|B'B^{(\Lambda)} \rangle ~,
\label{disc 2Lambda}
\end{equation}
where
$\hat{\Omega}=\omega(\hat{r}^2_\perp)+\omega((q-\hat{r})^2_\perp)$
and the states $\langle A'A^{(\Lambda)}|$, and
$|B'B^{(\Lambda)}\rangle $ are determined by the relations
\begin{equation}
\langle A'A^{(\Lambda)}|r_\perp\rangle
={\Phi}_{A'A}^{(\Lambda)}(r_\perp, r'_\perp)~, \;\;\langle
r_\perp|B'B^{(\Lambda)}\rangle ={\Phi}_{B'B}^{(\Lambda)}(-r_\perp,
-r'_\perp)~ .
\end{equation}
Note that, strictly speaking, in Eq.~(\ref{disc 2Lambda}) the use
of the equality sign is incorrect, because  $Y^2$-terms are
omitted, despite the fact that $Y$-terms with coefficients of the
same order in $g^2$ are kept. The $Y^2$-terms are omitted because
we want to compare the first two terms of the expansion in $Y$ on
both sides of the relation (\ref{boot el}). In the following we
also shall use the equality sign in this sense  for
discontinuities.

Let us turn to the contribution of intermediate states with three
jets (see Fig.~2b):
\begin{equation}
\frac{1}{-2\pi i}disc^{(3\Lambda)}_{s}{\cal A}_{AB}^{A'B'}=
-\frac{1}{2\pi}\sum_{ \tilde A  J \tilde B} {\cal A}_{AB}^{ \tilde
A J \tilde B}\; {\cal A}_{ \tilde A  J \tilde B}^{A'B'}d\rho_{
\tilde A  J \tilde B}~, \label{disc 3}
\end{equation}
where $ \tilde A $ and $ \tilde B$ are the jets produced in the
fragmentation regions of the particles $A$ and $B$, respectively,
and $J$ is the jet with total momentum $k$ produced in the
central region (it can be a single gluon, or two gluons, or a $q\bar
q$ pair). The amplitudes in Eq.~(\ref{disc 3}) have the form (\ref{A
mrk})). Passing to the scale $q_t$ at the PPR vertices we have
\[
{\cal A}^{ \tilde A J\tilde B}_{AB} = 2 s~ \Gamma^{a}_{ \tilde A
A}(q_t)
\frac{1}{r^2_{1}}\left(\frac{s_1}{q_{t}k_{t}}\right)^{\omega(r^2_1)}
\gamma^J_{ab}(r_1,
r_2)\frac{1}{{r^2_2}}\left(\frac{s_2}{k_{t}q_{t}}
\right)^{\omega(r^2_2)}\Gamma^{b}_{ \tilde B B}(q_t)~,
\]
\begin{equation}
{\cal A}_{ \tilde A J\tilde B}^{A'B'} = 2 s~ \Gamma^{a}_{A' \tilde
A }(q_t)
\frac{1}{r'^2_{1}}\left(\frac{s_1}{q_{t}k_{t}}\right)^{\omega(r'^2_1)}
\left(\gamma^J_{ab}(r'_1,
r'_2)\right)^*\frac{1}{{r'^2_2}}\left(\frac{s_2}{k_{t}q_{t}}
\right)^{\omega(r'^2_2)}\Gamma^{b}_{B' \tilde B }(q_t)~, \label{A
3jet}
\end{equation}
where $r_1=p_A-p_{ \tilde A }$, $r_2=p_{ \tilde B}-p_B$ and
$r'_{1,2}=q-r_{1,2}~.$  Note that $q=p_{A}-p_{A'}$ and that we can
put $q=q_\perp$. The Sudakov decomposition for the other momenta can be
written as
\begin{equation}
k=\beta p_1+\alpha p_2 +k_\perp~, \;\; s\alpha\beta
=k_t^2+k^2~,\;\; r_1=\beta p_1+q_{1\perp}~, \;\;r_2=-\alpha p_2
+q_{2\perp}~,
\end{equation}
so that $s_1={(k_t^2+k^2)}/{\beta}$ and $s_2=s\beta$.  The phase
space element has the form
\begin{equation}
 d\rho_{ \tilde A  J \tilde B}=d\phi_{ \tilde A }\frac{d\phi_{J}}
{2(2\pi)^{D-1}}d\phi_{ \tilde B}\frac{d^{D-2}r_{1\perp}}
{2s(2\pi)^{D-2}}d^{D-2}r_{2\perp}\frac{d\beta}{\beta}~, \label{rho
AJB}
\end{equation}
 and the limits  of integration
over $\beta$ are defined by the conditions $s_2\geq s_\Lambda~,
\;\;s_1\geq s_\Lambda~$. In the NLA we can put
\begin{equation}
\frac{k_t^2}{s_\Lambda}\geq \beta \geq \frac{s_\Lambda}{s}~,
\label{limits  el s}
\end{equation}
since an exact value of the limits of integration is important
only when the jet $J$ consists of a single gluon. Passing to the
integration variable $y=\ln \left(\beta s/(q_tk_t)\right)$,  we
have
\begin{equation}
{Y-y_{\Lambda}} \geq y \geq y_{\Lambda}~, \;\; y_{\Lambda}\equiv
\ln \left( \frac{s_\Lambda}{q_{t}k_t}\right)~.\label{limitsy el y}
\end{equation}

Again, in Eq.~(\ref{disc 3}) the sums are performed independently
for each jet.  The un-subtracted contribution to the colour octet
kernel from the production of real particles  is defined as
\begin{equation}
\langle r_{1\perp}|\hat{\cal K}_r^{(\Lambda)}|r_{2\perp}\rangle
={\cal K}_r^{(\Lambda)} (r_{1\perp},r_{2\perp};q_{\perp})=
\frac{f_{c_1c_2^{\prime }c}f_{c_{2}c_{2}^{\prime
}c}}{N_c(N_c^2-1)}\sum_{ J } \int \gamma_{c_{1}c_{2}}^{ J }\left(
q_{1},q_{2}\right) \left( \gamma_{c_{1}^{\prime }c_{2}^{\prime
}}^{ J} \left( r_{1}^{\prime},r_{2}^{\prime }\right) \right)
^{\ast } \frac{d\phi_{J}} {2(2\pi)^{D-1}}~.\label{real kernel un}
\end{equation}
Here and below the subscript $r$ denotes the contribution coming
from real particle production.

Since  the kernel depends on transverse momenta only,  the
dependence on $\beta$ in Eq.~(\ref{disc 3}) is contained only in
the phase space element (\ref{rho AJB}) and in the Regge factors
of the amplitudes (\ref{A 3jet}). In our approximation, the product
of these factors is reduced to
\begin{equation}
\left(\frac{s_1}{q_{t}k_{t}}\right)^{\omega(r^2_1)+\omega(r'^2_1)}
\left(\frac{s_2}{k_{t}q_{t}}
\right)^{\omega(r^2_2)+\omega(r'^2_2)}= 1+\Omega_1 (Y-y)+\Omega_2
y~, \label{Rfactor3el}
\end{equation}
where $\Omega_i=\omega(r^2_i)+\omega(r'^2_i)$. With NLO
accuracy we get
\begin{equation}
\int^{Y-y_{\Lambda}}_{y_{\Lambda}} dy\;(1+\Omega_1 (Y-y)+\Omega_2
y) = Y +(\Omega_1+\Omega_2)\frac{Y^2}{2}-y_{\Lambda}(1+\Omega_1
Y)-y_{\Lambda}(1+\Omega_2 Y)~. \label{int y}
\end{equation}
Omitting the irrelevant $Y^2$-terms  in Eq.~(\ref{int y}) we
obtain
\[
\frac{1}{-2\pi i}disc^{(3\Lambda)}_{s}{\cal A}_{AB}^{A'B'} =
\frac{2 sN_c}{(2\pi)^{D-1}}~ \int \frac{d^{D-2}r_1}{r^2_{1\perp}
r'^2_{1\perp}}\frac{d^{D-2}r_2}{r^2_{2\perp}
r'^2_{2\perp}}{\Phi}_{A'A}^{i(\Lambda)}(r_{1\perp},
r'_{1\perp}){\cal
K}_r^{(\Lambda)}(r_{1\perp},r_{2\perp};q_{\perp})
\]
\begin{equation}
\times \left[Y -y_{\Lambda}(1+\Omega_1 Y)-y_{\Lambda}(1+\Omega_2
Y)\right] {\Phi}_{B'B}^{i(\Lambda)} (-r_{2\perp}, -r'_{2\perp})~.
\label{disc3 un}
\end{equation}
Taking into account that the  terms with $y_{\Lambda}$  are
sub-leading, so that in the NLA they have to go together with the LO
impact factors and  kernel, we see that their contributions can be
combined with the terms of Eq.~(\ref{disc 2Lambda}) leading to
a subtraction in the
impact factors. This  subtraction is necessary in order to make the
impact factors independent on $s_\Lambda$. Note that, as usually,
any subtraction is ``scheme dependent". In fact, here we have fixed
already the scheme by the choice of the energy scale $q_t$. So we
define
\[
{\Phi}_{A'A}^{i}(r_\perp,
r'_\perp)={\Phi}_{A'A}^{i(\Lambda)}(r_\perp; r'_\perp) -\int
\frac{d^{D-2}r_1}{{r}_{1\perp}^{2}r'^{2}_{1\perp}}\Phi
_{A'A}^{i(B)}({r}_{1\perp},{r'}_{1\perp})
\]
\begin{equation}
\times {\cal K}_{r}^{(B)}({r}_{1\perp},{r}_{\perp}; q_\perp) \,
\ln \left( \frac{s_{\Lambda }} {({r}_1-{r})_t
q_t}\right)~,\label{impact}
\end{equation}
where $r'_{1\perp}=q_{\perp}-r_{1\perp}$, and the superscript $(B)$
refers to the Born approximation. In this approximation  the impact
factor and the kernel are given by Eqs.~(\ref{impact
un}) and (\ref{real kernel un}), respectively, where the jets consist of one
particle (for the kernel it is a gluon). Making use of the
hermiticity  property of the vertices $\left(\Gamma^i_{\tilde P P'}
\right)^*=\Gamma^i_{P'\tilde P} $ one can see from
Eqs.~(\ref{impact un}) and (\ref{impact}) that, apart from the
coefficient $i/\sqrt{N_c}$, our definition of
impact factors coincides with the one given in Ref.~~\cite{FF98} for
the case $s_0=q^2_t$.

Since we work in the LLA, NLO terms in the discontinuities can appear
only once. Therefore, with the
definition (\ref{impact}) the sum of Eqs.~(\ref{disc 2Lambda}) and
(\ref{disc3 un}) can be written as
\begin{equation}
\frac{1}{-2\pi i}disc^{(2\Lambda+3\Lambda)}_{s}{\cal
A}_{AB}^{A'B'}=  \frac{2 sN_c}{(2\pi)^{D-1}}\left(\langle
A'A|1+\hat{\Omega}Y|B'B\rangle  +\langle A'A^{(\Lambda)}|
\hat{\cal K}_r^{(\Lambda)} Y|B'B^{(\Lambda)}\rangle \right)~.
\label{disc 2+3}
\end{equation}

Consider now four jets in the unitarity relation.  Since the
leading contributions from such intermediate states contain  at
least $Y^2$, we need to take only the sub-leading piece,  coming from
the integration over rapidities of  jets  produced in the central
regions. Actually these jets can contain only one gluon each; the
amplitudes entering in the unitarity condition have the form
(\ref{A mrk}), with the Regge factors being omitted, and the vertices
being taken in the Born approximation. After the summation over the discrete
quantum numbers of the produced particles we recover in the discontinuity the
Born impact factors and kernels.  The integration over the rapidities
of the produced gluons with momenta $k_1$ and $k_2$ is performed by taking
into account the limitations
\begin{equation}
(p_{\tilde A}+k_1)^2=\frac{k_{1t}^2}{\beta_1}\geq s_\Lambda~,
\;\;(k_1+k_2)^2=\frac{\beta_1k_{2t}^2}{\beta_2}\geq s_\Lambda~,
\;\;(k_2+p_{\tilde B})^2=s\beta_2\geq s_\Lambda~.
\end{equation}
With the NLO accuracy the result of the integration is
\begin{equation}
\int
\frac{d\beta_1}{\beta_1}\frac{d\beta_2}{\beta_2}=\frac{Y^2}{2}- Y
(y_{1\Lambda}+y_{2\Lambda}+y_{3\Lambda})~,
\end{equation}
where
\begin{equation}
y_{1\Lambda}=\ln \left( \frac{s_{\Lambda }}
{q_tk_{1t}}\right)~,\;\;y_{2\Lambda}=\ln \left( \frac{s_{\Lambda
}} {k_{1t}k_{2t}}\right)~,\;\;y_{3\Lambda}=\ln \left(
\frac{s_{\Lambda }} {k_{2t}q_t}\right)~.
\end{equation}
The first term is irrelevant for us;  the terms with
$y_{i\Lambda}$  are necessary for subtractions in the second term
in Eq.~(\ref{disc 2+3}): here the first one and the last one serve for
subtractions in  the impact factors  of $A\rightarrow A'$ and
$B\rightarrow B'$ transitions, respectively, the second one in the
kernel.  After the subtraction the kernel becomes
\[
{\cal K}_{r}({q}_{1\perp},{_2}_{\perp}; q_\perp)={\cal
K}^{(\Lambda)}_{r} ({q}_{1\perp},{q}_{2\perp}; q_\perp)
 -\int
\frac{d^{D-2}r}{{r}_{\perp}^{2}(q-r)_\perp^{2}}{\cal
K}^{(B)}_{r}({q}_{1\perp},{r}_{\perp}; q_\perp)
\]
\begin{equation}
\times {\cal K}_{r}^{(B)}({r}_{\perp},{q}_{2\perp}; q_\perp) \,
\ln \left( \frac{s_{\Lambda }}
{({q}_1-{r})_t({q}_2-{r})_t}\right)~. \label{kernel r}
\end{equation}
It is worthwhile to note that the kernel is symmetric in its
first two arguments.

As a result we see that the discontinuity can be written as
\begin{equation}
\frac{1}{-2\pi i}disc_{s}{\cal A}_{AB}^{A'B'}= \frac{2
sN_c}{(2\pi)^{D-1}}\langle A'A|1+\hat{\cal K} Y |B'B\rangle ~,
\label{disc 2+3+4}
\end{equation}
where
\begin{equation}
\hat{\cal K}= \hat{\cal K}_r +\hat{\Omega}~\label{K=Kr+w}
\end{equation}
is the total colour octet kernel. As stated before, we keep
only the first two terms of the expansion in $Y$. Moreover, since we
work in the NLA, only leading and next-to-leading orders in the coefficients
of the expansion are under control.

\subsection{Discontinuities of one-gluon production amplitudes}

We now turn to the amplitude ${\cal A}_{AB}^{{A'}G{B'}}$ for the production
of a gluon $G$ with momentum $k$ in the MRK, for which we have
\[
k=q_1-q_2~,\;\; q_1=p_A-p_A'~,\;\; q_2=p_B'-p_B~,\;\;
\]
\begin{equation}
 k=\beta p_{1}+\alpha p_{2}+k_{\perp }~,\;\;
s\alpha\beta=-k_{\perp }^{2}={k}_t^{~2}~,\;\; \alpha  \ll 1 ~;
\;\; \beta \ll 1~.\;\;  \label{mrk gluon}
\end{equation}
Consequently we obtain
\begin{equation}
s_{1}\equiv
(p_{A'}+k)^2=s\alpha=\frac{k_t^2}{\beta}~,\;\;s_2\equiv(p_{B'}+k)^2
=s\beta~,\label{mrk gluon1}
\end{equation}
 and can put
\begin{equation}
q_1=\beta p_1+q_{1\perp }~,\;\; q_2=-\alpha p_2+q_{2\perp }~,\;\;
t_{1,2}=q_{1,2}^{2}= q_{1,2\perp }^{2}=-\vec{q}_{1,2}^{~2}~.
\label{mrk gluon2}
\end{equation}
The Reggeized form  of the production amplitude (\ref{A mrk}) now
reads
\begin{equation}
{\cal A}_{AB}^{{A'}G{B'}} = 2s\Gamma _{{A'}A}^{a} \left(
\frac{s_{1}} {{k}_{t}{q}_{1t}}\right)^{\omega (t_{1})}
\gamma_{ab}^{G}(q_{1},q_{2})\left(
\frac{s_{2}}{{k}_{t}{q}_{2t}}\right) ^{\omega (t_{2})} \Gamma
_{{B'}B}^{b}~. \label{A 3}
\end{equation}
The bootstrap relations (\ref{s1+s}) and (\ref{s2+s}) contain
discontinuities in  $s_1$, $s_2$ and $s$.  For brevity, we use the
term discontinuities, though actually we need and will calculate
only their imaginary parts. In analogy with the elastic case they
are calculated with the help of unitarity relations. In the
expressions for the discontinuities, given by unitarity relations
and calculated in NLA, only real parts of amplitudes are
important, so we can use the form (\ref{A mrk}). Similar to the
elastic case, in order to compare the left and right sides of the
relations (\ref{s1+s}) and (\ref{s2+s}) we need to use the same
scales of energies on both sides. The difference is that now we
have two independent energy variables (remember that $s_1s_2=s
k_t^2$) and two scales. Their choice is not unique; we prefer to
use variables and scales shown in Eq.~(\ref{A 3}), i.e. $s_1$ and
$s_2$ and the scales for them ${q}_{1t}{k}_{t}$ and
${k}_{t}{q}_{2t}$ respectively. We will use the notations
$Y_1=\ln\left(s_1/({q}_{1t}{k}_{t})\right)$ and
$Y_2=\ln\left(s_2/({k}_{t}{q}_{2t})\right)$ and, in analogy with
the elastic case, we will restrict ourselves to terms that are
linear in these variables.

\subsubsection*{$s_2$-channel discontinuity}

The discontinuity  can be found without large efforts, since its
calculation is very similar to the calculation of the
discontinuity of elastic amplitudes performed above. Indeed, in
Eq.~(\ref{A 3}) the vertex $\Gamma _{{A'}A}^{a}$ and the energy
factor of the $s_1$-channel are factorized, so that we may say we
need to calculate the discontinuity of the amplitude of the
process $R_1+B\rightarrow {G+{B'}}~,$ where the Reggeon $R_1$ with
momentum $q_1$ (see Fig.~3) plays the role of an incoming
particle.

\begin{figure}[htb]
\begin{center}
\begin{picture}(400,200)(-10,-100)

\LongArrow(40,-52)(40,52) \ZigZag(44,-29)(92,-29){3}{5}
\ArrowLine(92,-29)(96,-29) \GCirc(40,-29){4}{0}
\Text(34,62)[l]{$A^{\prime}$} \Text(34,-62)[l]{$A$}
\Text(55,-40)[l]{$q_1~~a$}

\CBoxc(100,0)(0.8,50){0}{0}
\Line(94,-19)(100,-13)
\Line(106,-19)(100,-13)
\Line(94,7)(100,13)
\Line(106,7)(100,13)
\LongArrow(100,33)(100,52)
\GCirc(100,-29){4}{0}
\GCirc(100,29){4}{0}
\ZigZag(104,29)(152,29){3}{5}
\ZigZag(104,-29)(152,-29){3}{5}
\ArrowLine(152,-29)(156,-29)
\ArrowLine(152,29)(156,29)
\Text(105,-10)[l]{$J$}
\Text(94,62)[l]{$G$}

\CBoxc(160,0)(0.8,50){0}{0}
\Line(154,-19)(160,-13)
\Line(166,-19)(160,-13)
\Line(154,7)(160,13)
\Line(166,7)(160,13)
\GCirc(160,-29){4}{0}
\GCirc(160,29){4}{0}
\ZigZag(164,29)(186,29){3}{3}
\ZigZag(164,-29)(186,-29){3}{3}
\ArrowLine(186,-29)(190,-29)
\ArrowLine(186,29)(190,29)
\Text(165,-10)[l]{$J_1$}
\Text(192,-29)[l]{$\cdots$}
\Text(192,29)[l]{$\cdots$}

\ZigZag(200,29)(222,29){3}{3}
\ZigZag(200,-29)(222,-29){3}{3}
\ArrowLine(222,-29)(226,-29)
\ArrowLine(222,29)(226,29)

\CBoxc(230,0)(0.8,50){0}{0}
\Line(224,-19)(230,-13)
\Line(236,-19)(230,-13)
\Line(224,7)(230,13)
\Line(236,7)(230,13)
\GCirc(230,-29){4}{0}
\GCirc(230,29){4}{0}
\ZigZag(234,29)(282,29){3}{5}
\ZigZag(234,-29)(282,-29){3}{5}
\ArrowLine(282,29)(286,29)
\ArrowLine(282,-29)(286,-29)
\Text(235,-10)[l]{$J_n$}

\GCirc(290,-29){4}{0}
\GCirc(290,29){4}{0}
\CBoxc(290,0)(0.8,50){0}{0}
\Line(284,-19)(290,-13)
\Line(296,-19)(290,-13)
\Line(284,7)(290,13)
\Line(296,7)(290,13)
\LongArrow(290,33)(290,52)
\ArrowLine(290,-52)(290,-33)
\Text(284,62)[l]{$B^{\prime}$}
\Text(284,-62)[l]{$B$}
\Text(308,-10)[r]{${\tilde{B}}$}

\DashLine(85,0)(305,0){4}

\end{picture}
\end{center}
\caption[]{Schematic representation of the $(n+2)$-jet
contribution to the $s_2$-channel discontinuity.}
\end{figure}
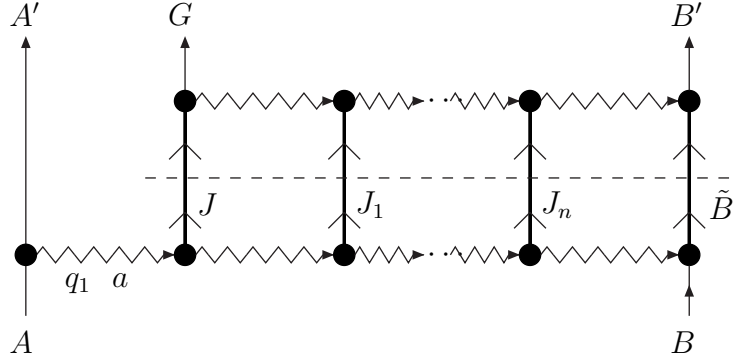

This discontinuity is calculated in the same way as in
Eq.~(\ref{disc 2+3+4}) with two  minor differences. The first of
them is evident:  the PPR vertex is replaced by the RPR vertex.
The second difference is related to the first one, but is not so
evident. It results from the change of energy scales. Remember
that when inserting our amplitudes into the unitarity relations we
have to pass from the 'natural' energy scale of these amplitudes
to 'external' scales.  With a change of scales both the scattering
vertices and the production vertices transform; their
transformation laws, however, are different.

In order to make this point clear (although, for the advanced reader, it
might be evident) let us consider the two-jet contribution in the $s_2$-channel
unitarity relation (see Fig.~3 with n=0)
\begin{equation}
\frac{1}{-2\pi i}disc^{(2\Lambda)}_{s_2}{\cal A}_{AB}^{A'GB'}=
-\frac{1}{2\pi}\sum_{ J \tilde B} {\cal A}_{AB}^{A'J
 \tilde B}\; {\cal A}_{ J  \tilde B}^{GB'}d\rho_{J  \tilde B}~.
 \label{disc s2 2}
\end{equation}
The amplitudes on the R.H.S. of this equation are defined
in Eq.~(\ref{A mrk}):
\[
{\cal A}^{ A' J  \tilde B}_{AB} = 2 s~ \Gamma^{a}_{ A'
A}\frac{1}{t_1}\left(\frac{s_{1}} {{q}_{1t}\tilde
 k_{t}}\right)^{\omega (t_{1})}
\gamma_{ab}^{J}(q_{1},r)\left(\frac{s_{2}}{{\tilde k_{t}
{r}_{t}}}\right) ^{\omega (r^{2})}\frac{1}{r^2} \Gamma^{b}_{
\tilde BB}~,\;\;
\]
\begin{equation}
{\cal A}_{J\tilde B}^{G B'} = \frac{2 s_2}{{r'^2}}~ \Gamma^{c}_{G
J}\left(\frac{s_2}{r'^{2}_t}\right)^{\omega(r'^2)}
\Gamma^{c}_{B'\tilde B}~, \label{3A 2jet}
\end{equation}
where $\tilde k=q_1-r$ is the $J$-jet momentum, and $r'=\tilde
k-k=q_2-r$; note that since transverse momenta are limited $\tilde
k$ has the same component along $p_1$ as $k~$, i.e. $\tilde
\beta=\beta$. Passing here to new energy scales, we obtain
\[
{\cal A}^{ A' J  \tilde B}_{AB} = 2 s~ \Gamma^{a}_{ A'
A}\frac{1}{t_1}\left( \frac{s_{1}}
{{q}_{1t}{k}_{t}}\right)^{\omega (t_{1})} \gamma_{ab}^{J}(q_{1},r;
k_t)\left( \frac{s_{2}}{{k}_{t}{q}_{2t}}\right)^{\omega
(r^{2})}\frac{1}{r^2} \Gamma^{b}_{ \tilde BB}(q_{2t})~,\;\;
\]
\begin{equation}
{\cal A}_{J\tilde B}^{G B'} = \frac{2 s_2}{{r'^2}}~ \Gamma^{c}_{G
J}(q_{2t})\left(\frac{s_2}{q^{2}_{2t}}\right)^{\omega(r'^2)}
\Gamma^{c}_{B'\tilde B}(q_{2t})~, \label{3A' 2jet}
\end{equation}
where the transformation of the PPR  vertices $\Gamma^{c}_{G J}$,
$\Gamma^{c}_{\tilde B B}$ and $\Gamma^{c}_{B'\tilde B}$, imposed
by the change of the energy scale, is defined in Eq.~(\ref{Gamma
scaled}), and
\begin{equation}
\gamma_{ab}^{J}(q_{1},r; k_t) =\left( \frac{{k}_{t}} {\tilde
{k}_{t}}\right)^{\omega (t_{1})}\gamma_{ab}^{J}(q_{1},r)\left(
\frac{{k}_{t}}{\tilde {k}_{t}}\right)^{\omega (r^{2})}~.
\label{gamma's scaled}
\end{equation}
Remember that, in order to write the formula in a compact way,
the perturbative expansion is not performed explicitly.
Actually the exponents in Eq.~(\ref{gamma's scaled}) should be expanded as in
Eq.~(\ref{Gamma scaled NLA}).

Now the difference with respect to the elastic case is clear.
Therefore we do not show a detailed treatment of three- and
four-jet contributions and only present the result, which in our
approximation  takes the form
\begin{equation}
\frac{1}{-2\pi i}disc_{s_2}{\cal A}_{AB}^{A'GB'}= \frac{2
sN_c}{(2\pi)^{D-1}}\Gamma _{{A'}A}\frac{1}{t_1}
<GR_1|1+\omega(t_{1})Y_1+\hat{\cal K}Y_2 |B'B>~. \label{disc RG
2+3+4}
\end{equation}
The state $<GR_1|$ describes the transition of the Reggeon
$R_1$ with momentum $q_1$ into the gluon $G$, and it is determined by the
equality
\[
<GR_1|r_\perp>_{ij} =~i\frac{f^{jbb'}}{N_c}\sum_{ J}\int \gamma^{J
}_{ib}(q_1, q_1- \tilde k_; k_t) \Gamma^{b'}_{G J }(q_{2t})
 d\phi_{ J }
\]
\begin{equation}
-\int
\frac{d^{D-2}r_1}{{r}_{1\perp}^{2}{r'}^{2}_{1\perp}}<GR_1^{(B)}|r_{1\perp}
>_{ij} {\cal K}_{r}^{(B)}({r}_{1\perp},{r}_{\perp}; q_{2\perp}) \, \ln
\left(\frac{s_{\Lambda }}{({r}_1-{r})_t k_t}\right)
~.\label{impact GR}
\end{equation}
Here  $i$ and $j$ are the   colour indices in the $t_1$ and $t_2$
channels, respectively,  $\tilde k$ is the $J$-jet momentum,
$r'_1=q_2-r_1$, and symmetrization with respect to
$r\leftrightarrow r'=q_2-r$ is understood.  Note that in
Eq.~(\ref{disc RG 2+3+4}) the total momentum of two $t$-channel
Reggeons (see Fig.~3) is $q_2=p_{B'}-p_B$, so that
$<r_{1\perp}|\hat{\cal K}|r_{2\perp}>={\cal
K}(r_{1\perp},r_{2\perp};q_{2\perp})$. The only feature of
Eq.~(\ref{impact GR}) that may require an explanation is the
argument of the logarithm in the subtraction term. It can be
understood easily. In the non-logarithmic terms the subtraction
comes from the three-particle intermediate state (see Fig.~3 with
n=1). The region of integration over rapidity of the jet $J$ with
momentum $\tilde k_1$ in the $s_2$-channel intermediate state is
limited by the conditions  $(\tilde k_1+p_{\tilde B})^2\geq
s_\Lambda~,\;\;(\tilde k+\tilde k_1)^2\geq s_\Lambda~,$ which in
the NLA (c.f. Eq.~(\ref{limits el s})) lead to:
\begin{equation}
 \frac{\tilde k_{1t}^2}{s_\Lambda} \geq\frac{\tilde{\beta_1}}
 {\beta} \geq \frac{s_\Lambda}{s_2}~.\;\;
 \label{limits inel s}
\end{equation}
For the integration variable $\tilde y=\ln \left(\tilde{\beta_1}
s/(q_{2t}\tilde k_{1t})\right)$  this means (c.f. Eq.~(\ref{limitsy
el y})):
\begin{equation}
{Y_2-\tilde y_{\Lambda}} \geq \tilde y \geq y_{2\Lambda}~, \;\;
\tilde y_{\Lambda}\equiv \ln \left( s_\Lambda/(k_{t}\tilde
k_{1t})\right)~,\;\; \tilde y_{2\Lambda}\equiv \ln \left(
s_\Lambda/(q_{2t}\tilde k_{1t})\right)~.\label{limits inel y}
\end{equation}
The term with $y_{2\Lambda}$ is used  for the subtraction in the
impact factor for the $B\rightarrow B'$ transition (c.f.
Eq.~(\ref{impact})). Therefore, the subtraction in $<GR_1|r_\perp>$
is performed by the term with $\tilde y_\Lambda$, and this explains the
argument of the logarithm in Eq.~(\ref{impact GR}).

\subsubsection*{$s$-channel discontinuity}

The calculation of the $s$-channel discontinuity is  more
intricate because it contains more components. Note that, since
$s'\equiv (p_{A'}+p_{B'})^2\simeq s$, we have to include here
discontinuities both properly in the $s$ channel (Fig.~4)

\begin{figure}[htb]
\begin{center}
\begin{picture}(400,400)(-30,-320)

\CBoxc(40,0)(0.8,50){0}{0}
\Line(34,-19)(40,-13)
\Line(46,-19)(40,-13)
\Line(34,7)(40,13)
\Line(46,7)(40,13)
\LongArrow(40,33)(40,52)
\GCirc(40,-29){4}{0}
\GCirc(40,29){4}{0}
\ArrowLine(40,-52)(40,-33)
\ZigZag(44,29)(92,29){3}{5}
\ZigZag(44,-29)(92,-29){3}{5}
\ArrowLine(92,-29)(96,-29)
\ArrowLine(92,29)(96,29)
\Text(34,-62)[l]{$A$}
\Text(22,-10)[l]{$\tilde{A}$}
\Text(34,62)[l]{$A^{\prime}$}

\CBoxc(100,0)(0.8,50){0}{0}
\Line(94,-19)(100,-13)
\Line(106,-19)(100,-13)
\Line(94,7)(100,13)
\Line(106,7)(100,13)
\GCirc(100,-29){4}{0}
\GCirc(100,29){4}{0}
\ZigZag(104,29)(126,29){3}{3}
\ZigZag(104,-29)(126,-29){3}{3}
\ArrowLine(126,-29)(130,-29)
\ArrowLine(126,29)(130,29)
\Text(105,-10)[l]{$J_1$}
\Text(132,-29)[l]{$\cdot \cdot$}
\Text(132,29)[l]{$\cdot \cdot$}

\ZigZag(140,29)(162,29){3}{3}
\ZigZag(140,-29)(196,-29){3}{8}
\ArrowLine(196,-29)(200,-29)
\ArrowLine(162,29)(166,29)
\GCirc(170,29){4}{0}
\ZigZag(174,29)(196,29){3}{3}
\ArrowLine(196,29)(200,29)
\LongArrow(170,33)(170,52)
\Text(164,62)[l]{$G$}
\Text(142,40)[l]{$r'_1~a'$}
\Text(173,40)[l]{$r'_2~b'$}
\Text(142,-40)[l]{$r_1~a$}
\Text(173,-40)[l]{$r_2~b$}

\Text(202,-29)[l]{$\cdot \cdot$}
\Text(202,29)[l]{$\cdot \cdot$}
\ZigZag(210,-29)(232,-29){3}{3}
\ArrowLine(232,-29)(236,-29)
\ZigZag(210,29)(232,29){3}{3}
\ArrowLine(232,29)(236,29)

\CBoxc(230,0)(0.8,50){0}{0}
\Line(224,-19)(230,-13)
\Line(236,-19)(230,-13)
\Line(224,7)(230,13)
\Line(236,7)(230,13)
\GCirc(230,-29){4}{0}
\GCirc(230,29){4}{0}
\ZigZag(234,29)(282,29){3}{5}
\ZigZag(234,-29)(282,-29){3}{5}
\ArrowLine(282,29)(286,29)
\ArrowLine(282,-29)(286,-29)
\Text(235,-10)[l]{$J_n$}

\GCirc(290,-29){4}{0}
\GCirc(290,29){4}{0}
\CBoxc(290,0)(0.8,50){0}{0}
\Line(284,-19)(290,-13)
\Line(296,-19)(290,-13)
\Line(284,7)(290,13)
\Line(296,7)(290,13)
\LongArrow(290,33)(290,52)
\ArrowLine(290,-52)(290,-33)
\Text(284,62)[l]{$B^{\prime}$}
\Text(284,-62)[l]{$B$}
\Text(308,-10)[r]{${\tilde{B}}$}

\DashLine(25,0)(305,0){4}

\Text(164,-84)[l]{$a$}

\CBoxc(40,-200)(0.8,50){0}{0}
\Line(34,-219)(40,-213)
\Line(46,-219)(40,-213)
\Line(34,-193)(40,-187)
\Line(46,-193)(40,-187)
\LongArrow(40,-167)(40,-148)
\GCirc(40,-229){4}{0}
\GCirc(40,-171){4}{0}
\ArrowLine(40,-252)(40,-233)
\ZigZag(44,-171)(92,-171){3}{5}
\ZigZag(44,-229)(92,-229){3}{5}
\ArrowLine(92,-229)(96,-229)
\ArrowLine(92,-171)(96,-171)
\Text(34,-262)[l]{$A$}
\Text(22,-210)[l]{$\tilde{A}$}
\Text(34,-138)[l]{$A^{\prime}$}

\CBoxc(100,-200)(0.8,50){0}{0}
\Line(94,-219)(100,-213)
\Line(106,-219)(100,-213)
\Line(94,-193)(100,-187)
\Line(106,-193)(100,-187)
\GCirc(100,-229){4}{0}
\GCirc(100,-171){4}{0}
\ZigZag(104,-171)(126,-171){3}{3}
\ZigZag(104,-229)(126,-229){3}{3}
\ArrowLine(126,-229)(130,-229)
\ArrowLine(126,-171)(130,-171)
\Text(105,-210)[l]{$J_1$}
\Text(132,-229)[l]{$\cdot \cdot$}
\Text(132,-171)[l]{$\cdot \cdot$}

\ZigZag(140,-171)(162,-171){3}{3}
\ZigZag(140,-229)(162,-229){3}{3}
\ArrowLine(162,-229)(166,-229)
\ArrowLine(162,-171)(166,-171)
\GCirc(170,-171){4}{0}
\GCirc(170,-229){4}{0}
\ZigZag(174,-171)(196,-171){3}{3}
\ZigZag(174,-229)(196,-229){3}{3}
\ArrowLine(196,-171)(200,-171)
\ArrowLine(196,-229)(200,-229)
\LongArrow(170,-167)(170,-148)
\Text(164,-138)[l]{$G$}
\ArrowLine(92,-171)(96,-171)
\ArrowLine(170,-200)(170,-175)
\ArrowLine(170,-225)(170,-200)
\Text(136,-160)[l]{$\tilde{r'_1}~~a'$}
\Text(179,-160)[l]{$\tilde{r'_2}~~b'$}
\Text(136,-240)[l]{$\tilde{r_1}~~a$}
\Text(179,-240)[l]{$\tilde{r_2}~~b$}
\Text(175,-210)[l]{$\tilde{G}$}

\Text(202,-229)[l]{$\cdot \cdot$}
\Text(202,-171)[l]{$\cdot \cdot$}
\ZigZag(210,-229)(232,-229){3}{3}
\ArrowLine(232,-229)(236,-229)
\ZigZag(210,-171)(232,-171){3}{3}
\ArrowLine(232,-171)(236,-171)

\CBoxc(230,-200)(0.8,50){0}{0}
\Line(224,-219)(230,-213)
\Line(236,-219)(230,-213)
\Line(224,-193)(230,-187)
\Line(236,-193)(230,-187)
\GCirc(230,-229){4}{0}
\GCirc(230,-171){4}{0}
\ZigZag(234,-171)(282,-171){3}{5}
\ZigZag(234,-229)(282,-229){3}{5}
\ArrowLine(282,-171)(286,-171)
\ArrowLine(282,-229)(286,-229)
\Text(235,-210)[l]{$J_n$}

\GCirc(290,-229){4}{0}
\GCirc(290,-171){4}{0}
\CBoxc(290,-200)(0.8,50){0}{0}
\Line(284,-219)(290,-213)
\Line(296,-219)(290,-213)
\Line(284,-193)(290,-187)
\Line(296,-193)(290,-187)
\LongArrow(290,-167)(290,-148)
\ArrowLine(290,-252)(290,-233)
\Text(284,-138)[l]{$B^{\prime}$}
\Text(284,-262)[l]{$B$}
\Text(308,-210)[r]{${\tilde{B}}$}

\DashLine(25,-200)(305,-200){4}

\Text(164,-284)[l]{$b$}

\end{picture}
\end{center}
\vspace{-1.cm}
\caption[]{Schematic representation of
contributions to the $s$-channel discontinuity: a) all produced
jets are far away in rapidity space from the gluon $G$; b) the
intermediate gluon $\tilde G$ is close to $G$.}
\end{figure}
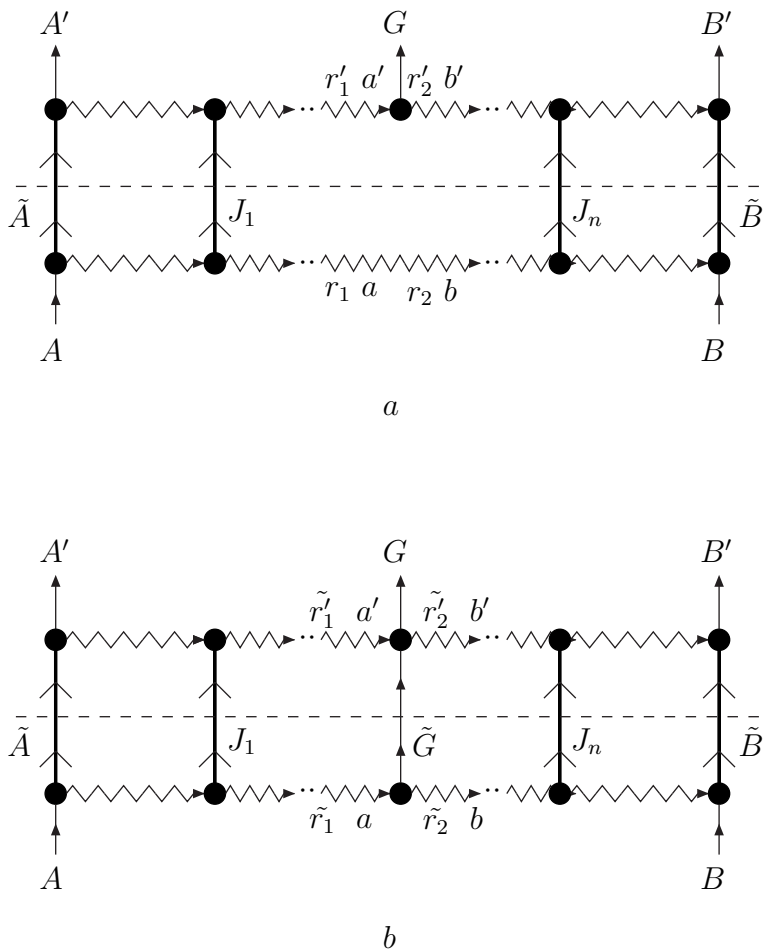
and in the $s'$-channel (Fig.~5).

We divide each of them  into two parts. The first one takes into
account contributions  of those intermediate jets which in
rapidity space are well separated from the gluon $G$.
Schematically this part is represented in Fig.~4a for the
$s$-channel, and in Fig.~5a for the $s'$-channel.
Here the momenta
$k_J$ of the produced jets are restricted by the condition $2kk_J
\geq s_\Lambda$.  In the second part, on the other hand, one of
produced jets in the rapidity space is close to the gluon $G$.
Actually in the NLA this jet consists of a single gluon. Its
momentum $k'$ is subject to the restriction $2kk' \leq s_\Lambda$.

\begin{figure}[htb]
\begin{center}
\begin{picture}(400,400)(-30,-320)

\CBoxc(40,0)(0.8,50){0}{0}
\Line(34,-19)(40,-13)
\Line(46,-19)(40,-13)
\Line(34,7)(40,13)
\Line(46,7)(40,13)
\LongArrow(40,33)(40,52)
\GCirc(40,-29){4}{0}
\GCirc(40,29){4}{0}
\ArrowLine(40,-52)(40,-33)
\ZigZag(44,29)(92,29){3}{5}
\ZigZag(44,-29)(92,-29){3}{5}
\ArrowLine(92,-29)(96,-29)
\ArrowLine(92,29)(96,29)
\Text(34,-62)[l]{$A$}
\Text(22,-10)[l]{$\tilde{A}$}
\Text(34,62)[l]{$A^{\prime}$}

\CBoxc(100,0)(0.8,50){0}{0}
\Line(94,-19)(100,-13)
\Line(106,-19)(100,-13)
\Line(94,7)(100,13)
\Line(106,7)(100,13)
\GCirc(100,-29){4}{0}
\GCirc(100,29){4}{0}
\ZigZag(104,29)(126,29){3}{3}
\ZigZag(104,-29)(126,-29){3}{3}
\ArrowLine(126,-29)(130,-29)
\ArrowLine(126,29)(130,29)
\Text(105,-10)[l]{$J_1$}
\Text(132,-29)[l]{$\cdot \cdot$}
\Text(132,29)[l]{$\cdot \cdot$}

\ZigZag(140,-29)(162,-29){3}{3}
\ZigZag(140,29)(196,29){3}{8}
\ArrowLine(196,29)(200,29)
\ArrowLine(162,-29)(166,-29)
\GCirc(170,-29){4}{0}
\ZigZag(174,-29)(196,-29){3}{3}
\ArrowLine(196,-29)(200,-29)
\LongArrow(170,37)(170,52)
\CArc(170,29)(8,90,270)
\ArrowLine(170,-25)(170,21)
\Text(164,62)[l]{$G$}
\Text(142,40)[l]{$r_1~a$}
\Text(173,40)[l]{$r_2~b$}
\Text(142,-40)[l]{$r'_1~a'$}
\Text(173,-40)[l]{$r'_2~b'$}

\Text(202,-29)[l]{$\cdot \cdot$}
\Text(202,29)[l]{$\cdot \cdot$}
\ZigZag(210,-29)(232,-29){3}{3}
\ArrowLine(232,-29)(236,-29)
\ZigZag(210,29)(232,29){3}{3}
\ArrowLine(232,29)(236,29)

\CBoxc(230,0)(0.8,50){0}{0}
\Line(224,-19)(230,-13)
\Line(236,-19)(230,-13)
\Line(224,7)(230,13)
\Line(236,7)(230,13)
\GCirc(230,-29){4}{0}
\GCirc(230,29){4}{0}
\ZigZag(234,29)(282,29){3}{5}
\ZigZag(234,-29)(282,-29){3}{5}
\ArrowLine(282,29)(286,29)
\ArrowLine(282,-29)(286,-29)
\Text(235,-10)[l]{$J_n$}

\GCirc(290,-29){4}{0}
\GCirc(290,29){4}{0}
\CBoxc(290,0)(0.8,50){0}{0}
\Line(284,-19)(290,-13)
\Line(296,-19)(290,-13)
\Line(284,7)(290,13)
\Line(296,7)(290,13)
\LongArrow(290,33)(290,52)
\ArrowLine(290,-52)(290,-33)
\Text(284,62)[l]{$B^{\prime}$}
\Text(284,-62)[l]{$B$}
\Text(308,-10)[r]{${\tilde{B}}$}

\DashLine(25,0)(305,0){4}

\Text(164,-84)[l]{$a$}

\CBoxc(40,-200)(0.8,50){0}{0}
\Line(34,-219)(40,-213)
\Line(46,-219)(40,-213)
\Line(34,-193)(40,-187)
\Line(46,-193)(40,-187)
\LongArrow(40,-167)(40,-148)
\GCirc(40,-229){4}{0}
\GCirc(40,-171){4}{0}
\ArrowLine(40,-252)(40,-233)
\ZigZag(44,-171)(92,-171){3}{5}
\ZigZag(44,-229)(92,-229){3}{5}
\ArrowLine(92,-229)(96,-229)
\ArrowLine(92,-171)(96,-171)
\Text(34,-262)[l]{$A$}
\Text(22,-210)[l]{$\tilde{A}$}
\Text(34,-138)[l]{$A^{\prime}$}

\CBoxc(100,-200)(0.8,50){0}{0}
\Line(94,-219)(100,-213)
\Line(106,-219)(100,-213)
\Line(94,-193)(100,-187)
\Line(106,-193)(100,-187)
\GCirc(100,-229){4}{0}
\GCirc(100,-171){4}{0}
\ZigZag(104,-171)(126,-171){3}{3}
\ZigZag(104,-229)(126,-229){3}{3}
\ArrowLine(126,-229)(130,-229)
\ArrowLine(126,-171)(130,-171)
\Text(105,-210)[l]{$J_1$}
\Text(132,-229)[l]{$\cdot \cdot$}
\Text(132,-171)[l]{$\cdot \cdot$}

\ZigZag(140,-171)(162,-171){3}{3}
\ZigZag(140,-229)(162,-229){3}{3} \ArrowLine(162,-229)(166,-229)
\ArrowLine(162,-171)(166,-171) \GCirc(170,-171){4}{0}
\GCirc(170,-229){4}{0} \ZigZag(174,-171)(196,-171){3}{3}
\ZigZag(174,-229)(196,-229){3}{3} \ArrowLine(196,-171)(200,-171)
\ArrowLine(196,-229)(200,-229)
\Text(164,-138)[l]{$G$} \ArrowLine(92,-171)(96,-171)
\ArrowLine(170,-200)(170,-175) \ArrowLine(170,-225)(170,-200)
\CArc(167,-171)(8,90,270) \LongArrow(167,-163)(167,-148)
\ArrowLine(167,-226)(167,-179)
\Text(136,-160)[l]{$\bar{r'_1}~~a'$}
\Text(179,-160)[l]{$\bar{r'_2}~~b'$}
\Text(136,-240)[l]{$\bar{r_1}~~a$}
\Text(179,-240)[l]{$\bar{r_2}~~b$} \Text(175,-210)[l]{$\bar{G}$}

\Text(202,-229)[l]{$\cdot \cdot$}
\Text(202,-171)[l]{$\cdot \cdot$}
\ZigZag(210,-229)(232,-229){3}{3}
\ArrowLine(232,-229)(236,-229)
\ZigZag(210,-171)(232,-171){3}{3}
\ArrowLine(232,-171)(236,-171)

\CBoxc(230,-200)(0.8,50){0}{0}
\Line(224,-219)(230,-213)
\Line(236,-219)(230,-213)
\Line(224,-193)(230,-187)
\Line(236,-193)(230,-187)
\GCirc(230,-229){4}{0}
\GCirc(230,-171){4}{0}
\ZigZag(234,-171)(282,-171){3}{5}
\ZigZag(234,-229)(282,-229){3}{5}
\ArrowLine(282,-171)(286,-171)
\ArrowLine(282,-229)(286,-229)
\Text(235,-210)[l]{$J_n$}

\GCirc(290,-229){4}{0}
\GCirc(290,-171){4}{0}
\CBoxc(290,-200)(0.8,50){0}{0}
\Line(284,-219)(290,-213)
\Line(296,-219)(290,-213)
\Line(284,-193)(290,-187)
\Line(296,-193)(290,-187)
\LongArrow(290,-167)(290,-148)
\ArrowLine(290,-252)(290,-233)
\Text(284,-138)[l]{$B^{\prime}$}
\Text(284,-262)[l]{$B$}
\Text(308,-210)[r]{${\tilde{B}}$}

\DashLine(25,-200)(305,-200){4}

\Text(164,-284)[l]{$b$}

\end{picture}
\end{center}
\vspace{-1.cm}
\caption[]{Schematic representation of
contributions to the $s'$-channel discontinuity: a) all produced
jets are far away in rapidity space from the gluon $G$; b) the
intermediate gluon $\bar G$ is close to $G$.}
\end{figure}
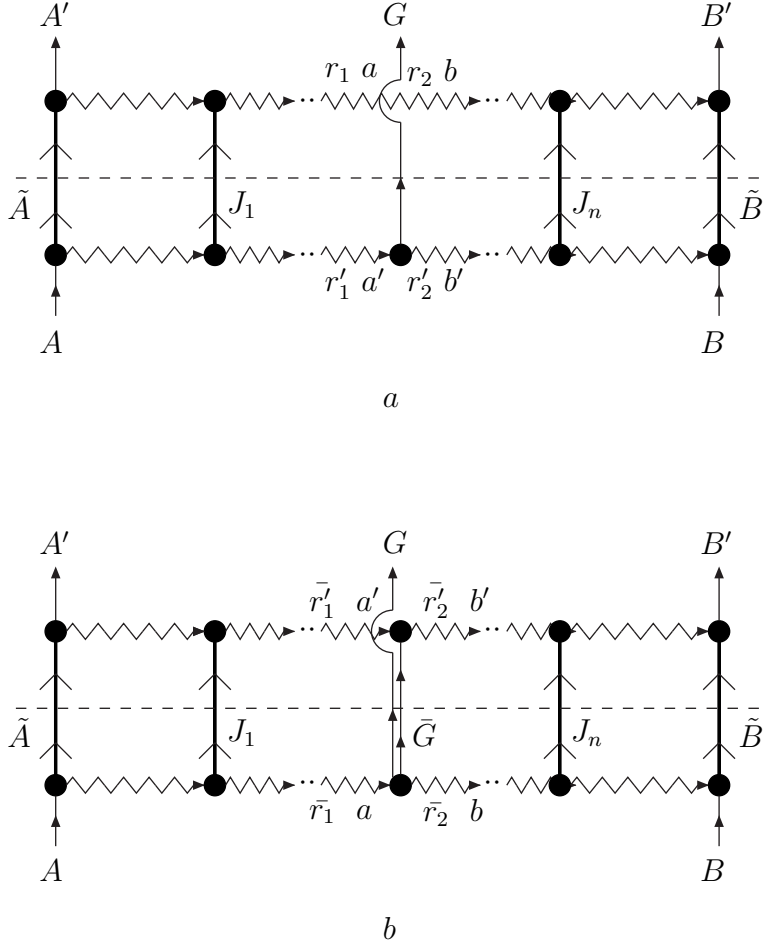

With the experience of the preceding calculations, instead of performing
a long sequence of derivations, we can immediately write down the result.
In our operator notation the terms of the discontinuity linear in $Y_{1,2}$
can be presented as
\begin{equation}
\frac{disc_{s}{{\cal A}_{AB}^{A'GB'}}}{-2\pi i}
=\frac{2sN_c}{(2\pi)^{D-1}}\;\langle{A^\prime A}|\hat{\cal
G}+Y_1{\hat{\cal K}}\;\hat{\cal G} +\hat{\cal G} {\hat{\cal
K}}Y_2|{B^\prime B}\rangle ~,\label{disc s}
\end{equation}
where $\hat{\cal G}$ is the operator of the gluon production. Note
that it changes the total two-Reggeon state momentum from $q_1$ to
$q_2$. With explicit reference to this fact the matrix element of
this operator takes the form
$$
\langle r_{1\perp}|\hat{\cal G}(q_1, q_2)|r_{2\perp}\rangle_{ij}
=\frac{f^{iaa'}f^{jbb'}}{N_c} \left[2
\gamma^{G}_{a'b'}(q_1-r_{1\perp},
q_2-r_{2\perp})\delta^{D-2}(r_{1\perp}-r_{2\perp})
 r_{1\perp}^{~2}\delta_{ab}\right.
$$
$$
\left.+\int_{\frac{{\tilde k}_t^{2}}{s_\Lambda}}^{1-\frac{
k_t^{2}}{s_\Lambda}} \frac{dx}{2x(1-x)} \sum_{ \tilde G }
 \frac{\gamma^{\tilde G(\tilde k)}_{ab }(\tilde r_1, \tilde r_2)}{(2\pi)^{D-1}}
\gamma_{a'b'}^{ \{G(k)\tilde G(-\tilde k) \}}\left( q_1-\tilde
r_{1},q_2-\tilde r_{2}\right)\right]
$$
$$
\left.+\int_{\frac{\bar{k}_t^{2}}{s_\Lambda}}^{1-\frac{
k_t^{2}}{s_\Lambda}} \frac{dx}{2x(1-x)} \sum_{ \bar G }
\gamma_{ab}^{\{G(k)\bar G(\bar k)\}}\left( \bar r_{1},\bar
r_{2}\right) \frac{ \gamma^{\bar G(-k')}_{a'b' }(q_1-\bar r_1,
q_2-\bar r_2)}{(2\pi)^{D-1}} \right.
$$
$$
-\int \frac{d^{D-2}r_\perp \langle r_{\perp}|\hat{\cal G}(q_1,q_2)
|r_{2\perp}\rangle_{ij}^{(B)}}{{r}^{2}_\perp
\left({r}-{q_1}\right)_\perp^{2}}  {\cal K}_{r}^{(B)}
\left(r_{1\perp}, {r}_\perp; q_{1\perp}\right) \, {\rm ln}\left(
\frac{s_{\Lambda }} {({r}-{r}_{1})_t  k_t}\right)
$$
\begin{equation} -\int \frac{d^{D-2}r_\perp \langle r_{1\perp}|\hat{\cal
G}(q_1,q_2) |r_{\perp} \rangle_{ij}^{(B)}}{{r}^{2}_\perp \left({r}
-{q_2}\right)_\perp^{2}}  {\cal K}_{r}^{(B)} \left( {r}_\perp,
{r}_{2\perp}; q_{2\perp}\right) \, {\rm ln}\left( \frac{s_{\Lambda
}} {( {r}-{r}_{2})_t k_t}\right) \label{RGR}~.
\end{equation}
The notations used here, apart from those of Eqs.~(\ref{mrk gluon}) -
(\ref{mrk gluon2}), are the following: $i$ and $j$ are the colour
indices in the $t_1$ and $t_2$ channels, $\tilde k$ and $\bar k$
are the momenta of the intermediate gluons $\tilde G$ and $\bar G$
in Fig.~4b and Fig.~5b, respectively, and $\tilde k=\tilde r_1-\tilde
r_2~,\;\;\bar k=\bar r_1-k-\bar r_2~,$ where $\tilde r_i$ and
$\bar r_i$ are the Reggeon momenta in these figures, with transverse
components being equal to $r_{i\perp},\;\; i=1,2$. Their Sudakov
decomposition is
\[
\tilde r_1= \frac{x\beta }{1-x}p_1+r_{1\perp}~,\;\;\tilde
r_2=\frac{\tilde k^2_\perp(1-x)} {x \beta s}p_2 +r_{2\perp}~,
\]
\begin{equation}
\bar r_1=\frac{\beta}{1-x}p_1+r_{1\perp}~,\;\;r_2=\left(\frac{\bar
k^2_\perp(1-x)} {x\beta s}+\frac{k^2_\perp}{\beta s}\right)p_2
+r_{2\perp}~\;\; .
\end{equation}
As always, the superscript $(B)$ in Eq.~(\ref{RGR}) refers to the leading
(Born) approximation; for $\langle r_{1\perp}|\hat{\cal
G}|r_{2\perp}\rangle$ it is given by the first term where the
gluon production vertex is taken in the LO.

Let us add a few necessary explanations. The first and the two
last terms on the R.H.S of Eq.~(\ref{RGR}) belong to the
contributions represented in Fig.~4a and  and Fig.~5a, which
evidently are equal (remember the signature). An interesting
aspect is that, unlike in Eq.~(\ref{impact GR}), in
Eq.~(\ref{RGR}) the gluon production vertex enters with its
natural scale. This can readily be seen from the two-jet
contributions. Indeed,  for example, the contribution to the
discontinuity from Fig.~4a at $n=0$ contains the product ${\cal
A}_{AB}^{ \tilde A
 \tilde B}\; {\cal A}_{ \tilde A  \tilde B}^{A'GB'}$.
Using the equalities $s=s_1s_2/k^2_t$ and $r_1=r_2\equiv r$ we can
rewrite the Regge factor in the amplitude ${\cal A}^{ \tilde A
\tilde B}_{AB}$  in the form
\begin{equation}
\left(\frac{s}{r^2_{t}}\right)^{\omega(r^2)}=
\left(\frac{s_1}{r_{1t}k_t}\right)^{\omega(r^2_1)}
\left(\frac{s_2}{r_{2t}k_t}\right)^{\omega(r^2_2)}~.
\end{equation}
Then, in the product ${\cal A}_{AB}^{ \tilde A
 \tilde B}\; {\cal A}_{ \tilde A  \tilde B}^{A'GB'}$,
 the Regge factors depending on $r_1$ and $r'_1=q_1-r_1$
 can be rewritten as
\begin{equation}
\left(\frac{s_1}{r'_{1t}k_t}\right)^{\omega(r'^2_1)}
\left(\frac{s_1}{r_{1t}k_t}\right)^{\omega(r^2_1)}=
\left(\frac{q_{1t}}{r'_{1t}}\right)^{\omega(r'^2_1)}
\left(\frac{q_{1t}}{r_{1t}}\right)^{\omega(r^2_1)}
\left(\frac{s_1}{q_{1t}k_t}\right)^{\omega(r'^2_1)+\omega(r^2_1)}~.
\label{regge factors on r1}
\end{equation}
The first two factors on the R.H.S. of this equation are used
for the transition (\ref{Gamma scaled}) to the scale $q_{1t}$ in
the Reggeon scattering vertices entering the  impact factor for
the $A\rightarrow A'$ transition (see (Eq.~\ref{impact un})).
After that in Eq.~(\ref{regge factors on r1}) we are left just with
the ``right" scale for $s_1$. The same procedure can be applied to
the Regge factors depending on $r_2$. Therefore there are no
additional factors which should be assigned to $\langle
r_{1\perp}|\hat{\cal G}(q_1,q_2)|r_{2\perp}\rangle$.

The second and third term in Eq.~(\ref{RGR}) correspond to Fig.~4b
and Fig.~5b,  respectively. Note that their contributions are
sub-leading, so that in Eq.~(\ref{disc s}) in the NLA they have to
go together with the LO impact factors and kernel.

Finally, the last two terms in Eq.~(\ref{RGR}) are subtraction
terms. As before, they appear as the result of the limits of
integration over rapidities of the produced jets (actually:
gluons).  Since these jets are separated in the rapidity space
from the gluon $G$, the full integration region for the
$s$-channel discontinuity is divided into two disconnected
subregions which actually are the integration regions of the
$s_1$- and $s_2$-channel discontinuities. Therefore we have two
subtraction terms. They are quite analogous to the subtraction
terms in the impact factors.

\subsection{Discontinuities of two-gluon production amplitudes}

The calculation of the discontinuities of the two-gluon production
amplitudes is performed quite in the same way as the one-gluon
case, so that we skip the description and only present the
results. Again, for brevity, we use the term discontinuities,
although, in reality, we calculate only their imaginary parts. We
use the notations of Section 2. The energy variables are
$s_i\equiv s_{i-1,i}$; their scales ${k}_{(i-1)t}{k}_{it}~,\;
\;k_{0t}\equiv q_{1t}~,\;\;k_{3t}\equiv q_{3t}~;\;$
$Y_i=\ln\left(s_i/({k}_{(i-1)t}{k}_{it})\right)~; \;$ in analogy
to the elastic case, only terms linear in $Y_i$ are kept.

For the $s_3$-channel discontinuity one obtains (cf. (\ref{disc RG
2+3+4}))
\begin{equation}
\frac{disc_{s_3}{\cal A}_{AB}^{A'G_1G_2B'}}{-2\pi i}= \frac{2
sN_c}{(2\pi)^{D-1}}\Gamma _{{A'}A}\frac{1}{t_1}\gamma^{G_1 }(q_1,
q_2)\frac{1}{t_2}
<G_2R_2|1+\omega(t_{1})Y_1+\omega(t_{2})Y_2+\hat{\cal K}Y_3
|B'B>~, \label{disc R2G2 2+3+4}
\end{equation}
where $<G_2R_2|r_\perp>$ is given by (\ref{impact GR}) with the
substitutions $G\rightarrow G_2~,\;\;R\rightarrow
R_2~,\;\;q_1\rightarrow q_2~,\;\;q_2\rightarrow q_3~; \;\;R_2$ is
the Reggeon with momentum $q_2$.

In the calculation of the $s_{13}$-channel discontinuity the
peculiarities of the calculations of both the $s_2$- and of the $s$-channel
discontinuities of the one-gluon production amplitudes are
combined. However, it does not require any new ideas, and the
calculation is straightforward. The result is
\begin{equation}
\frac{disc_{s_{13}}{\cal A}_{AB}^{A'G_1G_2B'}}{-2\pi i}= \frac{2
sN_c}{(2\pi)^{D-1}}\Gamma _{{A'}A}\frac{1}{t_1} <G_1R_1|\hat{\cal
G}_2+\omega(t_{1})Y_1\hat{\cal G}_2+Y_2{\hat{\cal K}}\;\hat{\cal
G}_2 +\hat{\cal G}_2 {\hat{\cal K}}Y_3|{B^\prime B}\rangle
~,\label{disc s13}
\end{equation}
where $<G_1R_1|r_\perp>$ is given by (\ref{impact GR}) with the
substitution $G\rightarrow G_1~; \;\;\hat{\cal G}_2$ is the
operator for the production of the gluon $G_2$; it changes the
total momentum of the two-Reggeon state from $q_2$ to $q_3$, so
that  the matrix elements of this operator are given by
(\ref{RGR}) with the substitutions $k\rightarrow
k_2~,\;\;q_1\rightarrow q_2~,\;\;q_2\rightarrow q_3~$.

The calculation of the $s$-channel discontinuity resembles the
one-gluon case and gives
\begin{equation}
\frac{disc_{s}{{\cal A}_{AB}^{A'G_1G_2B'}}}{2\pi i}
=\frac{2sN_c}{(2\pi)^{D-1}}\;\langle{A^\prime A}|\hat{\cal
G}_1\hat{\cal G}_2+Y_1{\hat{\cal K}}\;\hat{\cal G}_1\hat{\cal G}_2
+\hat{\cal G}_1Y_2{\hat{\cal K}}\;\hat{\cal G}_2+\hat{\cal
G}_1\hat{\cal G}_2 {\hat{\cal K}}Y_3|{B^\prime B}\rangle
~.\label{disc-two s}
\end{equation}
The discontinuities in the channels $s_1$ and $s_{02}$ are
obtained from (\ref{disc R2G2 2+3+4}) and (\ref{disc s13}),
respectively, by suitable replacements. The $s_2$-channel
discontinuity is
\begin{equation}
\frac{disc_{s_{2}}{\cal A}_{AB}^{A'G_1G_2B'}}{-2\pi i}= \frac{2
sN_c}{(2\pi)^{D-1}}\Gamma _{{A'}A}\frac{1}{t_1} \langle
G_1R_1|1+\omega(t_{1})Y_1+ {\hat{\cal
K}}Y_2+\omega(t_{3})Y_3|G_2R_2\rangle \frac{1}{t_3}\Gamma
_{{B'}B}~.\label{disc-two s2}
\end{equation}

\section{Bootstrap conditions}

Use of Eq.~(\ref{A mrk}) at $n=0$ on the R.H.S. of the relation
(\ref{boot el}) gives an explicit form of the bootstrap condition
for the elastic amplitudes:
\begin{equation}
\frac{1}{-2\pi i} disc_{s} {\cal A}_{AB}^{A'B'}
=\frac{\omega(t)}{2}\;\frac{2s}{t} \;\Gamma^i_{A^\prime
A}\left(\frac{s}{q_t^2} \right)^ {\omega(t)}\; \Gamma^i_{B^\prime
B}~. \label{boot el s}
\end{equation}
Taking into account Eq.~(\ref{disc 2+3+4}) and comparing
non-logarithmic terms in both sides of this equation, we obtain
\begin{equation}
\frac{2sN_c}{(2\pi)^{D-1}}\langle {A^\prime A}|{B B^\prime
}\rangle =\frac{\omega(t)}{2}\;\frac{2s}{t}\;\Gamma_{A^\prime A}\;
\Gamma_{B^\prime B}~.\label{con el 0}
\end{equation}
The terms proportional to $Y$ give
\begin{equation}
\frac{2sN_c}{(2\pi)^{D-1}}\langle {A^\prime A}|\hat {\cal K}|{B
B^\prime }\rangle
=\frac{\left(\omega(t)\right)^2}{2}\;\frac{2s}{t}\;\Gamma_{A^\prime
A}\; \Gamma_{B^\prime B}~. \label{con el 1}
\end{equation}
Instead of using the last equation it is more convenient to
consider the difference between (\ref{con el 1}) and (\ref{con el 0}),
the latter being multiplied by $\omega (t)$; this leads to
\begin{equation} \langle {A^\prime A}|\hat {\cal
K}-\omega(t)|{B B^\prime }\rangle=0~.\label{con el k}
\end{equation}
In the LO these equalities follow from  so called strong bootstrap
conditions for the impact factors and for the kernel:
\begin{equation}
\langle {A'A^{(B)}}| = \frac{g}{2}\Gamma^{(B)}_{A'A} \langle
R_\omega^{(B)}|~,\;\;\;|{BB'^{(B)}}\rangle =
|R_\omega^{(B)}\rangle \frac{g}{2}\Gamma^{(B)}_{B'B}~,\;\;\;
\biggl( \hat {\cal K}^{(B)} - \omega^{(1)}(t) \biggr)
|R_\omega^{(B)}\rangle = 0~,\;\;\;\label{strong LO}
\end{equation}
where the superscripts $(B)$ and $(1)$ mean Born and one-loop
approximations, respectively, and $|R^{(B)}_{\omega}\rangle $ is
the universal (process independent) eigenfunction of the kernel
with the eigenvalue $\omega^{(1)}(t)$. The normalization of
$|R^{(B)}_{\omega}\rangle $ is determined by (\ref{con el 0}):
\begin{equation}
\frac{g^2N_c t}{2(2\pi)^{D-1}}\langle
R_\omega^{(B)}|R_\omega^{(B)}\rangle =
\omega^{(1)}(t)~.\label{norm RB}
\end{equation}
The fulfilment of Eq.~(\ref{strong LO}) is known from
Ref.~\cite{BFKL}. Moreover, it is known that
\begin{equation}
\langle  r_\perp|R_\omega^{(B)}\rangle=1~.\label{R LO}
\end{equation}
The conditions which, with the account of Eq.~(\ref{strong LO}),
follow from Eqs.~(\ref{con el 0}) and
(\ref{con el 1}) in the NLO are the following:
\begin{equation}
\frac{gN_ct}{(2\pi)^{D-1}}<A'A^{(1)}|R_\omega^{(B)}>
=\omega^{(1)}(t)
\Gamma^{(1)}_{A'A}+\frac{\omega^{(2)}(t)}{2}\Gamma^{(B)}_{A'A}~
\label{impact 1 condition}
\end{equation}
and
\begin{equation}
\frac{g^2{N_c}t}{2(2\pi)^{D-1}}<R_\omega^{(B)}|\hat{\cal
K}^{(1)}|R_\omega^{(B)}> =\omega^{(1)}(t)\omega^{(2)}(t)~,
\label{kernel 1 condition}
\end{equation}
where the superscript $(2)$ indicates the two-loop contribution.  It is
just the NLO bootstrap conditions for the octet impact factors and
kernel derived in Ref.~\cite{FF98}.

Let us now turn to the bootstrap relation (\ref{s2+s}) for
one-gluon production amplitudes. Using Eq.~(\ref{A mrk}) at $n=1$
we obtain its explicit form:
\[
\Re\;\left[\frac{1}{-2\pi i} \left(disc_{s_2}+disc_{s}\right)
{\cal A}_{AB}^{A'GB'} \right]=
\]
\begin{equation}
\frac{\omega(t_2)}{2}\;2s \;\Gamma^i_{A^\prime
A}\frac{1}{t_1}\left(\frac{s_1}{q_{1t}k_t} \right)^
{\omega(t_1)}\;\gamma^G_{ij}(q_1,q_2)\;\frac{1}{t_2}\left(\frac{s_2}{k_tq_{2t}}\right)^
{\omega(t_2)}\; \Gamma^j_{B^\prime B}~. \label{boot prod s2+s}
\end{equation}
Comparing here non-logarithmic terms with the account of
Eqs.~(\ref{disc 2+3+4}) and (\ref{disc s}) we have
\begin{equation}
\frac{2sN_c}{(2\pi)^{D-1}}\;\left[\langle{A^\prime A}| \hat{\cal
G}|{B B^\prime}\rangle +\Gamma_{A^\prime A}\frac{1}{t_1} \;
\langle {GR_1}|{B B^\prime}\rangle\;\right]
=\frac{\omega(t_2)}{2}\;{2s}\Gamma^i_{A^\prime
A}\frac{1}{t_1}\gamma^G_{ij}(q_1,q_2)\frac{1}{t_2}
\Gamma^j_{B^\prime B}~.\label{bootnl AGB}
\end{equation}
Comparison of  terms proportional to $Y_1$ and $Y_2$ in (\ref{boot
prod s2+s}) gives
\[
\frac{2sN_c}{(2\pi)^{D-1}}\;\left[\langle{A^\prime A}| {\hat{\cal
K}}\;\hat{\cal G}|{B B^\prime}\rangle +\Gamma_{A^\prime
A}\frac{\omega(t_1)}{t_1} \; \langle {GR}|{B
B^\prime}\rangle\;\right]
\]
\begin{equation}
=\frac{\omega(t_2)}{2}\;{2s}\Gamma^i_{A^\prime
A}\frac{\omega(t_1)}{t_1} \;\gamma^G_{ij}(q_1,q_2)\frac{1}{t_2}
\Gamma^j_{B^\prime B}~\label{booty1 AGB}
\end{equation}
and
\[
\frac{2sN_c}{(2\pi)^{D-1}}\;\left[\langle{A^\prime A}| \hat{\cal
G}{\hat{\cal K}}|{B B^\prime}\rangle +\Gamma_{A^\prime
A}\frac{1}{t_1} \; \langle {GR}|{\hat{\cal K}}|{B
B^\prime}\rangle\;\right]\;
\]
\begin{equation}
=\frac{\omega(t_2)}{2}\;{2s}\Gamma^i_{A^\prime A}\frac{1}{t_1}
\;\gamma^G_{ij}(q_1,q_2)\frac{ \omega(t_2)}{t_2}
\Gamma^j_{B^\prime B}\;\;, \label{booty2 AGB}
\end{equation}
respectively.
Subtracting Eq.~(\ref{bootnl AGB}) (multiplied by $\omega(t_1)$) from
Eq.~(\ref{booty1 AGB}) we obtain
\begin{equation}
\langle{A^\prime A}|\left({\hat{\cal K}}-\omega(t_1)\right)
\hat{\cal G}|{B B^\prime}\rangle =0~.\label{bootk AGB}
\end{equation}
Since $\langle r_\perp|\hat{\cal G}|{B B^\prime}\rangle$ even in the Born
approximation is a complicated function of $r_\perp$ depending on
$q_2$ and the gluon momentum $k$, it can be considered as an
arbitrary function. Therefore, in order to satisfy Eq.~(\ref{bootk AGB})
the equality
\begin{equation}
\langle{A^\prime A}|\left({\hat{\cal K}}-\omega(t_1)\right)
 =0~\label{bootkl AGB}
\end{equation}
must be fulfilled in the NLO. It is easy to see that the difference
between Eq.~(\ref{booty2 AGB}) and
Eq.~(\ref{bootnl AGB}) (multiplied by $\omega(t_2)$) is zero, assuming that
\begin{equation}
\left({\hat{\cal K}}-\omega(t_1)\right)|{B B^\prime}\rangle
 =0~. \label{bootkr AGB}
\end{equation}
Actually Eq.~(\ref{bootkr AGB}) is equivalent to Eq.~(\ref{bootkl AGB}),
so that we do not obtain a new relation.

Eq.~(\ref{bootkl AGB}) requires  the fulfilment, in the NLO, of the
strong bootstrap conditions for the impact factors and for the kernel:
\begin{equation}
\langle {A'A}| = \frac{g}{2}\Gamma_{A'A} \langle
R_\omega|~,\;\;\;|{BB'}\rangle = |R_\omega\rangle
\frac{g}{2}\Gamma_{B'B}~,\;\;\; \biggl( \hat {\cal K} - \omega(t)
\biggr) |R_\omega \rangle = 0~.\;\;\;\label{strong NLO}
\end{equation}
Note that the normalization of $|R_\omega \rangle$ is fixed if we
take into account (\ref{norm RB}) and (\ref{impact 1 condition}):
\begin{equation}
\frac{g^2t{N_c}}{2(2\pi)^{D-1}} \langle R_\omega|R_\omega \rangle
={\omega(t)} ~.\label{norm R}
\end{equation}

So the first important consequence derived from the bootstrap
relations for one-gluon production is the strong form (\ref{strong
NLO}), (\ref{norm R}) of the bootstrap conditions for impact
factors and for the kernel in the NLO. Moreover, these relations
give a new restriction on the Reggeon vertices and on the gluon
trajectory.

In the leading order, using the equalities (\ref{strong LO})
and (\ref{norm RB})  (writing explicitly the total momenta of the
two-Reggeon states, in order to avoid uncertainties, and denoting by $R_i$
the Reggeons with momenta $q_i$), we obtain from (\ref{bootnl AGB})
\begin{equation}
\frac{gt_1}{2}\langle R^{(B)}_{\omega}(q_1)|\hat{\cal
G}^{(B)}|R^{(B)}_\omega (q_2)\rangle_{ij} + \langle
GR_1^{(B)}|R^{(B)}_\omega(q_2) \rangle_{ij}=
\gamma^{G(B)}_{ij}(q_1,q_2) \frac{g}{2}\langle
R^{(B)}_\omega(q_2)|R^{(B)}_\omega(q_2) \rangle~. \label{boot born
GR}
\end{equation}
This equality is a particular case of the stronger version:
\begin{equation}
\frac{gt_1}{2}\langle R^{(B)}_{\omega}(q_1)|\hat{\cal G}^{(B)} +
\langle GR_1^{(B)}|= \gamma^{G(B)}(q_1,q_2) \frac{g}{2}\langle
R^{(B)}_\omega (q_2)|~.  \label{boot strong born GR}
\end{equation}
Indeed, Eq.~(\ref{boot born GR}) can be obtained from
Eq.~(\ref{boot strong born GR}) by projection on the state
$|R^{(B)}_\omega \rangle$. The fulfillment of Eq.~(\ref{boot strong
born GR}) was demonstrated long ago~\cite{BLF}, since it has been used
in the proof of the gluon Reggeization in the LLA. Together with
Eq.~(\ref{bootnl AGB}) it gives a {new} NLO bootstrap
condition, which can be written as
\begin{equation}
\frac{gN_ct_2}{(2\pi)^{D-1}}\left[\frac{gt_1}{2}\langle
R_{\omega}(q_1)|\hat{\cal G}| R_{\omega}(q_2)\rangle + \langle
GR_1| R_{\omega}(q_2)\rangle\right] =
\omega(t_2)\gamma^G_{ij}(q_1,q_2)~. \label{boot RG}
\end{equation}

Note that this condition has a "weak" form (it is a condition for
matrix elements, not for state vectors). In this sense it is
analogous to the conditions for the impact factors and for the
kernel, obtained from the elastic bootstrap. Since the bootstrap
relations for one-gluon production  lead to the "strong" form
(\ref{strong NLO}) of the conditions for the impact factors and
for the kernel, it is natural to expect that (\ref{boot born GR})
can be strengthened  to the form
\begin{equation}
\frac{gt_1}{2}\langle R_{\omega}(q_1)|\hat{\cal G} + \langle
GR_1|= \gamma^{G}(q_1,q_2) \frac{g}{2}\langle R_\omega (q_2)|~,
\label{boot strong  GR}
\end{equation}
which is an evident generalization of (\ref{boot born GR}) to the
NLO, by consideration of the bootstrap relations for two-gluon
production amplitudes. Moreover, one could expect that the
bootstrap relations for the two-gluon production will give us a
new restriction on the Reggeon vertices and trajectory. It turns
out that the first expectation is correct; instead, the second
one, fortunately, is not justified.

Indeed, let us turn to the relations (\ref{s23+s13+s}) and
(\ref{s01+s02+s}) and consider the first of them, substituting the
discontinuities (\ref{disc R2G2 2+3+4})-(\ref{disc-two s}) in the
L.H.S. of the relation, and the Reggeized form (\ref{A mrk}) of
the two-gluon production amplitude on the R.H.S. Comparing
non-logarithmic terms on both sides of the bootstrap relation we
obtain, with the help of (\ref{strong NLO}),
\[
\frac{gN_ct_3}{(2\pi)^{D-1}}\left[\frac{gt_1t_2}{2}\langle
R_{\omega}(q_1)|\hat{\cal G}_1\hat{\cal G}_2|R_\omega(q_3) \rangle
+ t_2\langle G_1R_1|\hat{\cal G}_2|R_\omega(q_3)
\rangle+\gamma^{G_1}(q_1,q_2)\langle G_2R_2|R_\omega(q_3)
\rangle\right]
\]
\begin{equation}
=\omega(t_3)
\gamma^{G_1}(q_1,q_2)\gamma^{G_2}(q_2,q_3)~.\label{boot 2-0}
\end{equation}
The terms proportional to $Y_2$ give
\[
\frac{gN_ct_3}{(2\pi)^{D-1}}\left[\frac{gt_1t_2}{2}\langle
R_{\omega}(q_1)|\hat{\cal G}_1\hat{\cal K}\hat{\cal
G}_2|R_\omega(q_3) \rangle + t_2\langle G_1R_1|\hat{\cal
K}\hat{\cal G}_2|R_\omega(q_3)
\rangle+\omega(t_2)\gamma^{G_1}(q_1,q_2)\langle
G_2R_2|R_\omega(q_3) \rangle\right]
\]
\begin{equation}
=\omega(t_2)\omega(t_3)
\gamma^{G_1}(q_1,q_2)\gamma^{G_2}(q_2,q_3)~. \label{boot 2-Y2}
\end{equation}
It is easy to see that a comparison of the terms proportional to $Y_1$
and $Y_3$ does not lead to a new condition. Together with
(\ref{strong NLO}) the corresponding equations are reduced to
(\ref{boot 2-0}).

Subtracting (\ref{boot 2-0}) (multiplied by $\omega(t_2)$) from
(\ref{boot 2-Y2}) we obtain
\begin{equation}
\left(\frac{gt_1}{2}\langle R_{\omega}(q_1)|\hat{\cal G}_1 +
\langle G_1R_1|\right) \left(\hat{\cal
K}-\omega(t_2)\right)\hat{\cal G}_2|R_\omega(q_3) \rangle
 =0~,\label{boot 2-Y2-0}
\end{equation}
which means that $(gt_1/{2})\langle R_{\omega}(q_1)|\hat{\cal G}_1 +
\langle G_1R_1|$ is the eigenvector of the kernel with the
eigenvalue $\omega(t_2)$, i.e. it must be proportional $\langle
R_{\omega}(q_2)|$. Taking into account (\ref{boot RG}) we arrive at
(\ref{boot strong GR}).

One can easily see that a double use of this bootstrap condition,
together with the normalization of $|R_\omega \rangle$, guarantees
that (\ref{boot 2-0}) is fulfilled. Moreover, it is not difficult to
see that the relations (\ref{s01+s02+s}) do not lead to new conditions.

\section{Summary}
The phenomenon of gluon Reggeization, which is very important for
high energy QCD, has been proven in the leading logarithmic
approximation, but it still remains a hypothesis in the
next-to-leading approximation.  The requirement of compatibility
of the gluon Reggeization with  $s$-channel unitarity imposes
stringent restrictions on  the gluon Regge trajectory and on the
vertices of Reggeon interactions. The restrictions deduced from
elastic scattering amplitudes in next-to-leading order were
derived several years ago~\cite{FF98}. They are known as the
bootstrap conditions for the color octet impact factors and for
the BFKL kernel, and they were proven to be satisfied. Moreover,
subsequently it was shown (see~\cite{FP02} and references therein)
that the stronger conditions on the impact factors and kernel are
also fulfilled.

In this paper we have considered restrictions on the Reggeon
vertices and trajectory, which emerge from amplitudes of gluon
production in  multi-Regge kinematics. We have shown that the
requirement of compatibility of the multi-Regge form of these
amplitudes with  $s$-channel unitarity leads, in particular, to
the strong bootstrap  conditions on the colour octet impact
factors and on the kernel that we have mentioned above. Besides
this, a new bootstrap condition has been derived. The most urgent
problem now is a proof of fulfillment of this new condition. It
will provide the possibility to prove the hypothesis of gluon
Reggeization in the NLA. Indeed, the bootstrap conditions are
extraordinarily significant. The proof of the gluon Reggeization
in the leading logarithmic approximation was constructed just on
the basis of these conditions. An analogous proof can be
constructed in the next-to-leading approximation as
well~\cite{tbp}.

\vspace{0.2cm} \noindent \underline{\bf Acknowledgments:}  V.S.F.
thanks the Alexander von Humboldt foundation for the research
award, the Universit\"at Hamburg and DESY, the Dipartimento di
Fisica dell'Universit\`a della Calabria and the Istituto Nazionale
di Fisica Nucleare - Gruppo collegato di Cosenza for their warm
hospitality while a part of this work was done.

\end{document}